\PassOptionsToPackage{table}{xcolor}
 \documentclass[useAMS,usenatbib,twocolumn]{aastex631}

 \usepackage{amsmath}
 \usepackage{amssymb}
 \usepackage{natbib}
 \usepackage{multirow}
 \usepackage{graphicx}\graphicspath{{graphics/}}
 \usepackage{color}
 \usepackage{epstopdf}

\definecolor{blackberry}{HTML}{8D1D75}

\received{\today}
\revised{\today}
\accepted{\today}
\submitjournal{ApJ}

\shorttitle{Galaxy Quenching at High Redshifts}
\shortauthors{Kimmig et al.}

\begin{document}

\title{Blowing out the Candle: How to Quench Galaxies at High Redshift -- an Ensemble of Rapid Starbursts, AGN Feedback and Environment}

\correspondingauthor{Lucas C. Kimmig}
\email{lkimmig@usm.lmu.de}
\author{Lucas C. Kimmig}
\affil{Universit\"ats-Sternwarte M\"unchen, Fakult\"at f\"ur Physik, Ludwig-Maximilians-Universit\"at, Scheinerstr.\ 1, D-81679 M\"unchen, Germany}
\author{Rhea-Silvia Remus}
\affil{Universit\"ats-Sternwarte M\"unchen, Fakult\"at f\"ur Physik, Ludwig-Maximilians-Universit\"at, Scheinerstr.\ 1, D-81679 M\"unchen, Germany}
\author{Benjamin Seidel}
\affil{Universit\"ats-Sternwarte M\"unchen, Fakult\"at f\"ur Physik, Ludwig-Maximilians-Universit\"at, Scheinerstr.\ 1, D-81679 M\"unchen, Germany}
\author{Lucas M. Valenzuela}
\affil{Universit\"ats-Sternwarte M\"unchen, Fakult\"at f\"ur Physik, Ludwig-Maximilians-Universit\"at, Scheinerstr.\ 1, D-81679 M\"unchen, Germany}
\author{Klaus Dolag}
\affil{Universit\"ats-Sternwarte M\"unchen, Fakult\"at f\"ur Physik, Ludwig-Maximilians-Universit\"at, Scheinerstr.\ 1, D-81679 M\"unchen, Germany}
\affil{Max-Planck-Institute for Astrophysics, Karl-Schwarzschild-Str.\ 1, D-85748 Garching, Germany}
\author{Andreas Burkert}
\affil{Universit\"ats-Sternwarte M\"unchen, Fakult\"at f\"ur Physik, Ludwig-Maximilians-Universit\"at, Scheinerstr.\ 1, D-81679 M\"unchen, Germany}
\affil{Max-Planck-Institute for Extraterrestrial Physics, Giessenbacherstr. 1, 85748 Garching, Germany}
\affil{Excellence Cluster ORIGINS, Boltzmannstrasse 2, 85748 Garching, Germany}

\begin{abstract}
Recent observations with JWST and ALMA have revealed extremely massive quiescent galaxies at redshifts of $z=3$ and higher, indicating both rapid onset and quenching of star formation. Using the cosmological simulation suite Magneticum Pathfinder we reproduce the observed number densities and stellar masses, with 36 quenched galaxies of stellar mass larger than $3\times10^{10}M_\odot$ at $z=3.42$. We find that these galaxies are quenched through a rapid burst of star-formation and subsequent AGN feedback caused by a particularly isotropic collapse of surrounding gas, occurring on timescales of around 200Myr or shorter. The resulting quenched galaxies host stellar components which are kinematically fast rotating and alpha-enhanced, while exhibiting a steeper metallicity and flatter age gradient compared to galaxies of similar stellar mass. The gas of the galaxies has been metal enriched and ejected. We find that quenched galaxies do not inhabit the densest nodes, but rather sit in local underdensities. We analyze observable metrics to predict future quenching at high redshifts, finding that on shorter timescales $<500\,$Myr the ratio $M_\mathrm{bh}/M_*$ is the best predictor, followed by the burstiness of the preceding star-formation, $t_{50}-t_{90}$ (time to go from $50\%$ to $90\%$ stellar mass). On longer timescales, $>1$Gyr, the environment becomes the strongest predictor, followed by $t_{50}-t_{90}$, indicating that at high redshifts the consumption of old and lack of new gas are more relevant for long-term prevention of star-formation than the presence of a massive AGN. We predict that relics of such high-z quenched galaxies should best be characterized by a strong alpha enhancement.
\end{abstract}

\keywords{galaxies: general -- high-redshift -- formation -- evolution -- methods: numerical}

\section{Introduction}
In the last few years, the high redshift universe beyond $z=2$ has become increasingly accessible to observations due to instruments like ALMA and JWST. The detection of galaxies already massive at redshifts beyond $z=6$ \citep[e.g.,][]{finkelstein:2015,oesch:2018,bouwens:2020,adams:2023,curtislake:2023,donan:2023,labbe:2023} has challenged our current understanding of structure formation, as models and simulations struggle to reproduce and predict the properties of such galaxies at high redshifts \citep[e.g.][]{harikane:2022,kakimoto23}.
This has led to the question whether these galaxies could feasibly exist within the framework of $\Lambda$CDM, as the measured masses were larger than what models predicted to be allowed \citep[e.g.,][]{behroozi:2018,tacchella:2022_too_massive,boylan:2023,lovell:2023}. However, these early JWST observations were usually obtained using photometry, and spectroscopic follow-ups have found several of them to reside at lower redshifts \citep{harikane:2023_spec} or to be less massive than originally assumed \citep[e.g.,][]{desprez:2023}. Furthermore, high-redshift properties may be particularly challenging for current generation spectral energy distribution (SED) codes \citep{wang:2023}.

Beyond the existence of particularly massive galaxies, the detection of quiescent galaxies at high redshift provides another challenge to our understanding of galaxy and structure formation \citep[e.g.,][]{nanayakkara:2022,carnall:2023,long:2023,carnall:2023onequenchie,glazebrook:2023}.
At low redshifts, several different mechanisms to quench a galaxy are known: Most prominently, mergers between massive galaxies involving little gas are a common cause for quenching \citep[e.g.,][]{barnes:1988,barnes:1992,barnes:1996,bois:2011,dekel:2006etgs}. In such cases, the feedback from central supermassive black holes (SMBH), so called active galactic nucleus (AGN) feedback, is thought to be an important contributor to the quenching \citep[e.g.,][]{combes:2017,man:2018}. However, also stellar feedback has been shown to be efficient in supressing star formation \citep[e.g.,][]{feldmann:2015}.
Another established quenching mechanism for galaxies beyond $M_\mathrm{vir}\approx10^{12}M_\odot$ is mass quenching, as a galaxy that reaches such large halo masses inhabits a hot gas halo that cannot be penetrated by cold inflowing streams, cutting the galaxy off from the continuous inflow of cold gas from the cosmic web \citep{birnboim:2003,dekel:2006}. Finally, the hot gas environment of galaxy groups and particularly galaxy clusters prohibit cold gas supply to their member galaxies, independent of the galaxies' mass, and thus all galaxies in such environments eventually run out of gas and quench. This family of processes acting in a cluster environment is known as environmental quenching \citep[e.g.,][]{gabor:2015,zinger:2016,lotz:2019}.

However, all of these processes work on timescales larger than those available to the quenched galaxies found when the Universe was not even $1\,$Gyr old. Even in the quenched galaxy population observed at low redshifts, some galaxies are found to have quenched at very early times, while for many others the quenching occurred on longer timescales more recently \citep{carnall:2018,tacchella:2022}. So the question arises how such galaxies can effectively quench at such early times.

Observations report evidence for the quenching at cosmic noon to be rapid in nature, especially before $z\approx2$, preceded by a rapid and massive starburst \citep{carnall:2018,tacchella:2023_threegal,kakimoto23}. The star formation timescales of these galaxies are found to be on the order of $150-300\,$Myr. For example, \citet{belli:2023} report a galaxy at $z\approx2.5$ with a starburst time of less than $300\,$Myr, while 
\citet{carnall:2023onequenchie} find a quenched galaxy at $z=4.658$ with a stellar mass of $M_*\approx3.8\times10^{10}M_\odot$ which was formed in a starburst event lasting around $200\,$Myr. At low redshifts, such a massive starburst followed by rapid quenching is usually only known for Post-Starburst (PSB) galaxies \citep[e.g.,][]{wild:2016}, caused either by rapid AGN quenching due to merging in case of field-PSBs, or by the cluster environment \citep[e.g.,][]{lotz:2021}. Overall, observations indicate that rapid quenching is increasingly more common at higher redshifts than at lower redshifts \citep{belli:2019,park:2023}, most likely caused by efficient gas funneling to the galaxy centers.

Evidence for a co-evolution of starbursts and AGN feedback through cosmic time has been reported from the GAMA/DEVILS survey by \citet{dsilva:2023}, with the star formation rate density and the AGN density rising until $z\approx2$, before declining again. Similarly, \citet{tozzi:2023} found a co-evolution between star formation and black hole accretion activity, albeit not all galaxies that show AGN activity are found to be quenched, indicating that AGN feedback does not always lead to immediate quenching of the star formation in a galaxy. Similarly, in their review on black holes at cosmic noon, \citet{fan:2023} also report on the co-evolution of star formation and black hole properties.

Furthermore, while usually strong outflows are associated with AGN driven quasars \citep[e.g.,][]{vayner:2023}, there are also outflows detected in galaxies without direct evidence for an AGN to be responsible \citep{wylezalek:2020}. This implies either that the outflows originate from a different source, or that the outflows are more long lasting than the AGN activity itself. Similarly, neutral outflows are reported by \citet{belli:2023} for a rapidly quenching galaxy at $z=2.445$ which exhibits neither X-ray nor radio emissions, and only gas emission lines might indicate an AGN as the source of these outflows. 
In fact, JWST observations by \citet{veilleux:2023} recently showed detailed imaging of an active AGN outflow at $z=1.6$ that burns a channel into the surrounding medium. This indicates that it is not the mechanical feedback that is impacting the galaxy but rather the combination of radiative feedback and heating from the outflow which could contribute to the lack of gas in the galaxy. Nevertheless, it is possible that the gas could also have already been driven out previously.
A complete picture of the interplay between star formation, AGN feedback, and quenching at high redshifts has yet to be established. Nonetheless, it is becoming apparent that AGN activity and massive rapid starbursts are coupled and connected to the process that quenches galaxies at high redshifts.

Simulations and models that have been able to reproduce the observed overall star formation rate density have found particularly the number density of quenched galaxies to depend on the included AGN feedback. In the family of IllustrisTNG and MilleniumTNG models, MilleniumTNG is large enough to successfully reproduce star formation rate densities and number densities of high redshift galaxies \citep{kannan:2023}. However, the quenching solely depends on the onset of the AGN feedback \citep{park:2023}. As shown by \citet{hartley:2023} and \citet{kurinchi:2023}, as soon as the kinetic feedback kicks in the galaxies are quenched. This directly correlates the possible quenching times with the timescales implemented for the onset of AGN feedback, which vary strongly between simulations. In this framework, however, the observed extremely early quenching times found for example by \citet{nanayakkara:2022}, \citet{glazebrook:2023}, or \citet{carnall:2023,carnall:2023onequenchie} cannot successfully be reproduced, as shown by \citet{hartley:2023}.
Furthermore, \citet{tollet:2019} found for the NIHAO simulations that stellar feedback from supernovae can generate outflows that reach beyond the virial radius, and therefore can have non-negligible impact on the feedback energy budget. That quenching can potentially also occur due to stellar feedback alone was also found by \citet{merlin:2012}.
Nevertheless, feedback from AGN plays an important role in the quenching of galaxies at cosmic noon: as shown by \citet{lovell:2023b} for the FLARES zoom simulations, comparing runs with and without AGN feedback, quenching does not occur if the AGN feedback is not included.

On the modeling side, \citet{dsilva:2023_flares_sharks} compared high-redshift quenching from the cosmological simulation FLARES with predictions from the semi-analytic model SHARKS, and found the onset of quenching to be earlier in FLARES than in SHARKS due to stronger feedback. \citet{lagos:2023} showed that SHARK 2.0, with a new AGN model that accounts for angular momentum transfer to the black hole spin as well as angular momentum transfer from the intergalactic medium to stars, can now successfully reproduce the observed fractions of quenched galaxies. Using the semi-analytic model on the Uchuu simulations, \citet{houston:2023} reported a trend for galaxies with masses below $\log(M_*/M_\odot)=9$ and above $\log(M_*/M_\odot)=11.5$ to spend the longest overall time as quenched galaxies, with the low mass galaxies switching star formation on and off, while the high mass galaxies tend to quench more permanently. This is similar to the results by \citet{merlin:2012} who also found massive galaxies to be quenched by short powerful starburst, while small mass galaxies switch on and off due to small starbursts. The latter was also reported by \citet{dome:2023} for the small mass haloes in IllustrisTNG.

To disentangle the contributions of stellar feedback from the starbursts and the SMBHs, but also address the question of the importance of the environment on quenching of galaxies at high redshifts, we study the processes that lead to the quenching of galaxies at high redshifts using the Magneticum\footnote{www.magneticum.org} pathfinder simulation suite, reviewed shortly in Sec.~\ref{sec:sim}. In Sec.~\ref{sec:stats}, we analyze the global properties of the simulated galaxies in comparison to high-redshift observations, demonstrating that the simulation is well suited for this study. Furthermore, the sample of quenched galaxies is introduced, where for more details on the overall properties of the quenched galaxy sample we refer the reader to the companion paper of this study by Remus \& Kimmig (submitted). Sec.~\ref{sec:quench} then analyzes in detail the processes that lead to quenching, including quenching timescales and angular momentum properties. We focus on the impact of the environment for the quenching of galaxies in Sec.~\ref{sec:env}. Finally, we discuss possible observable indicators to predict quenching on a longer timescale, disentangling the relative importance of star formation, AGN feedback, and environment on the quenching at high redshifts in Sec.~\ref{sec:disc}, before we conclude our paper in Sec.~\ref{sec:conc}. As mentioned above, this paper is the companion paper to a second study by Remus \& Kimmig (submitted), hereafter RK23, where we study the properties of the quenched galaxies and their future evolution and rejuvenation in more detail.

\section{The Magneticum Simulation}\label{sec:sim}
As we will show in Sec.~\ref{subsec:sample}, a simulation requires both a large volume as well as a high particle resolution to sufficiently reproduce the observed massive quiescent galaxies at high redshifts. To this end, we employ Box3~uhr of the hydrodynamical cosmological simulation suite Magneticum Pathfinder (www.magneticum.org), which has a volume of $(128Mpc/h)^3$ and particle masses of $m_\mathrm{dm}=3.6\times10^{7} M_{\odot}/h$ and $m_\mathrm{gas} = 7.3\times10^{6} M_{\odot}/h$, for dark matter and gas, respectively. As every gas particle can spawn up to four stellar particles, their mass resolution is $m_\mathrm{*}\approx1/4^{\rm th}m_\mathrm{gas}\approx1.8\times10^{6} M_{\odot}/h$, such that galaxies of $M_*\approx 3\times10^{10}M_\odot$ are well resolved. The softening lengths for every component are $\epsilon_\mathrm{dm} = \epsilon_\mathrm{gas} = 1.4~\mathrm{kpc}/h$ and $\epsilon_\mathrm{*} = 0.7~\mathrm{kpc}/h$. The assumed cosmology is that of WMAP-7, with $h=0.704$, $\Omega_m = 0.272$, $\Omega_b = 0.0451$, $\Omega_\lambda = 0.728$, $\sigma_8 = 0.809$ and $n_s = 0.963$ as given by \citet{komatsu:2011}. 

The code employed for the simulation is a modified version of GADGET-2 \citep{springel:2005}, with upgrades to the implementation of the smoothed particle hydrodynamics (SPH), covering thermal conduction \citep{dolag:2004} and artificial viscosity \citep{dolag:2005}, as well as other improvements \citep{donnert:2013,beck:2015}. The included baryonic physics are described by \citet{teklu:2015}, with star formation and stellar feedback based on \citet{springel:2003}. Included are feedback and metal enrichment from supernovae type Ia and II as well as from stars on the asymptotic giant branch (AGB) based on \citet{tornatore:2004,tornatore:2007} and \citet{wiersma:2009}, with upgrades to the metal enrichment described by \citet{dolag:2017}. Black hole seeding, evolution and feedback are implemented via the model from \citet{steinborn:2015}, based on the work by \citet{fabjan:2010}. The simulations include background radiation from UV/X-rays as well as the cosmic microwave background as given by \citet{haardt:2001}.

Identification of galaxies is done via the structure finder SUBFIND \citep{springel:2001, dolag:2009}. The most massive galaxy within a given halo is termed the 'central', with all other galaxies present within the halo being 'satellite' galaxies. The tracing of galaxies through time is performed via L-BaseTree \citep{springel:2005b}, allowing us to track the individual assembly history of galaxies in our study down to $z=2$ where the simulation Box3~uhr stopped. The range up to $z\approx15$ is covered via nine snapshots with full particle data in spacing of $\approx400\,\mathrm{Myr}$, with one in particular at $z=3.42$. This is comparable to the average redshift of observed quenched galaxies from \citet{nanayakkara:2022} or \citet{long:2023}.

\section{Galaxies at high redshifts}\label{sec:stats}
\begin{figure*}
  \begin{center}
    \includegraphics[width=.95\columnwidth]{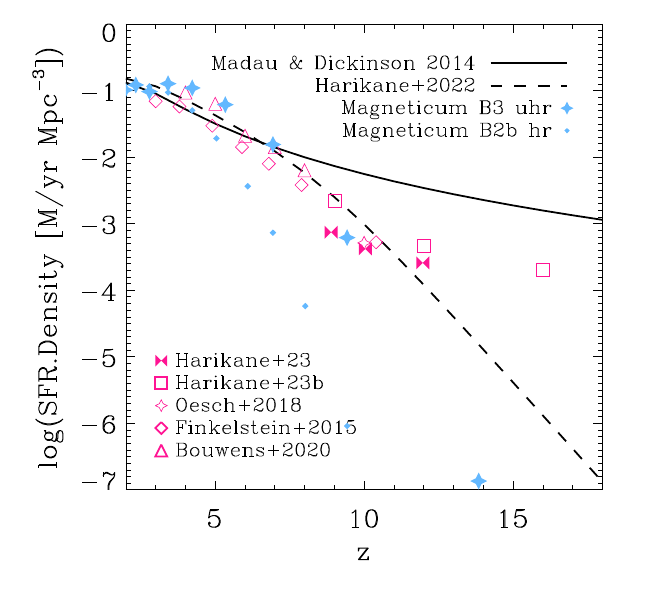}
    \includegraphics[width=.95\columnwidth]{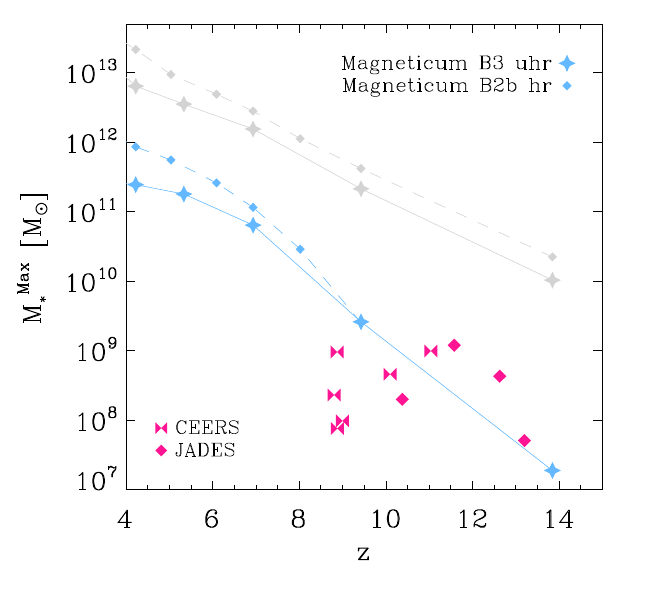}
    \caption{\textit{Left:} Star formation rate density as a function of redshift for Magneticum Box3~uhr (large blue stars) and Box2b~hr (small blue stars). Model predictions from \citet{madau:2014} and \citet{harikane:2022} are included as solid and dashed lines, respectively. Observational data from different methods are included: Spectroscopic data from JWST by \citet{harikane:2023_spec} is shown as solid pink bow-ties, while photometric JWST data is included as open pink squares \citep{harikane:2023_phot}. The UV-based measurement from HST by \citet{oesch:2018} at $z\approx10$ is marked as open pink star. For a larger redshift range, ALMA-based measurement by \citet{bouwens:2020} and UV-based measurements sampled over different instruments by \citet{finkelstein:2015} are included as open pink triangles and diamonds, respectively. We clearly see that the higher resolved simulation reproduces the star formation rate density up to much larger redshifts than the lower resolution simulation.  
    \textit{Right:} Maximum stellar mass of a galaxy within the simulation at a given redshift, for Box3~uhr (large blue stars and solid line) and Box2b~hr (small blue stars and dashed line). Observed galaxies with spectroscopic redshifts from the JWST programs CEERS and Jades, taken from \citet{harikane:2023_spec}, are included as pink bow-ties and diamonds, respectively. The gray lines show the maximum dark matter mass at these redshifts in Box3~uhr (solid gray) and Box2b~hr (dashed gray).
    }
  {\label{fig:props}}
\end{center}
\end{figure*}
The detection of galaxies at high redshifts had already been done in the last few years by HST, and for those that contain large gas fractions also by ALMA, but with the advent of JWST we have now reached the ability to also statistically probe the properties of galaxies at high redshifts in unprecedented detail. However, simulations still struggle to reproduce and understand many of the observed properties of galaxies at cosmic dawn, as this requires large simulation volumes while simultaneously also providing high resolutions. This is particularly challenging, and only few simulations that are currently available are capable of fulfilling this criterion. Thus, we will introduce the global properties of the galaxies found in our simulation in more detail in this section.

\subsection{Galaxy Sample from z=15 to z=2}\label{subsec:sample}
The ability of a simulation to answer the questions of how and why massive quiescents formed and quenched is closely related to its size and resolution. To demonstrate this, the left panel of Fig.~\ref{fig:props} shows the star formation rate density as a function of redshift for two simulations of the Magneticum Suite, the larger but less resolved Box2b~hr (small blue stars, see \citeauthor{lotz:2019,kimmig:2023} for more details on this simulation), and the better resolved but smaller Box3~uhr that is used throughout this study (large blue stars). 

We find that at redshifts below $z<4$ both simulations agree with the observational results by \citet{finkelstein:2015} and \citet{bouwens:2020}, as well as with the relations found by \citet{madau:2014} and \citet{harikane:2022}. At higher redshifts than $z=4$, however, the star formation rate density of Box2b~hr noticeably falls below that of Box3~uhr and the observed values. The star formation rate density of the higher resolved Box3~uhr still agrees well with the observations from different instruments up to $z\approx10$, including UV-based observations by \citet{oesch:2018} from HST and \citet{finkelstein:2015} from different instruments, ALMA observations from \citet{bouwens:2020}, and JWST observations from photometry \citep{harikane:2023_phot} and spectroscopy \citep{harikane:2023_spec}. It also agrees with the fit by \citet{harikane:2022}, however, it does not agree with the prediction from \citet{madau:2014}, which also diverts from the observed results as already discussed by \citet{harikane:2022} and \citet{harikane:2023_phot}. Unfortunately, at even higher redshifts the resolution of our simulation becomes too small such that we cannot statistically probe galaxies beyond $z\approx10$. We furthermore caution the use of the smallest mass bin due to the stellar resolution, and include it primarily to get an impression of the expectations at this lowest mass range.

When instead we consider the maximum stellar mass as a function of redshift, as shown in the right panel of Fig.~\ref{fig:props}, we find that the larger simulation volume of Box2b~hr generally contains more massive galaxies at a given redshift, which is caused by the bigger box containing more massive nodes of the cosmic web. This can also be seen when looking at the total maximum mass at a given redshift for both simulations, which are included as gray lines in the right panel of Fig.~\ref{fig:props}. However, while we find constantly larger values in total mass for the larger volume simulation, we again see the effect of resolution on the baryonic component at higher redshifts. At about $z\approx8$ the lower resolution box produces no galaxies with larger stellar masses than those in the higher resolution volume, while for $z>8$ there are no galaxies which contain the minimum number of $20$~stellar particles identified in the box. For the higher resolution volume, we find that galaxies are already present with masses above $M_*>10^7M_\odot$ at $z\approx14$. Nevertheless, if the simulation with the larger simulation volume would have better resolution, it would also contain more massive galaxies at high redshifts as the halo masses are accordingly larger. This mirrors the results by \citet{remus:2023}, who discuss the consequences of box volume on finding massive structures at high redshift in more detail. It is also in good agreement with the results found for the MilleniumTNG and IllustrisTNG Simulations by \citet{kannan:2023}, who have the same problems in obtaining massive galaxies at high redshifts, albeit they get closer to observations at $z\approx12$ as their box volume is large and their resolution relatively high.

Overall, we see from the right panel of Fig.~\ref{fig:props} that the massive galaxies observed at particularly high redshifts of $z>8$ are generally covered well by the predictions from the Box3~uhr simulation, albeit some observed galaxies are more massive than those predicted by the simulation at $z\approx12$. Note that we here only include galaxies that have been spectroscopically confirmed, and not those that are currently debated from photometric redshift to challenge $\Lambda$CDM. 
As discussed above, we expect these to be reproducible within the standard framework of $\Lambda$CDM for an even larger simulation volume on the given resolution level. Nevertheless, such a simulation would prove highly computationally expensive and consequently does not exist to our knowledge.
We conclude that our simulation Box3~uhr successfully reproduces the star formation rate density and hosts galaxies of comparable stellar masses to those observed at high redshifts, providing us with a galaxy sample that can be used to study the evolution and formation of galaxies at cosmic dawn.

\subsection{Quenched Galaxies: Selection and Statistics}
The recent observations of very massive quiescent galaxies at redshifts of $z\approx3-4$ by for example \citet{nanayakkara:2022} and \citet{long:2023} raise crucial questions to how they could have formed and why they quenched. However, before we can address these particular questions, we must define quiescence and compare the overall quenched fractions found at different redshifts to observations.
\begin{figure}
  \begin{center}
    \includegraphics[width=\columnwidth]{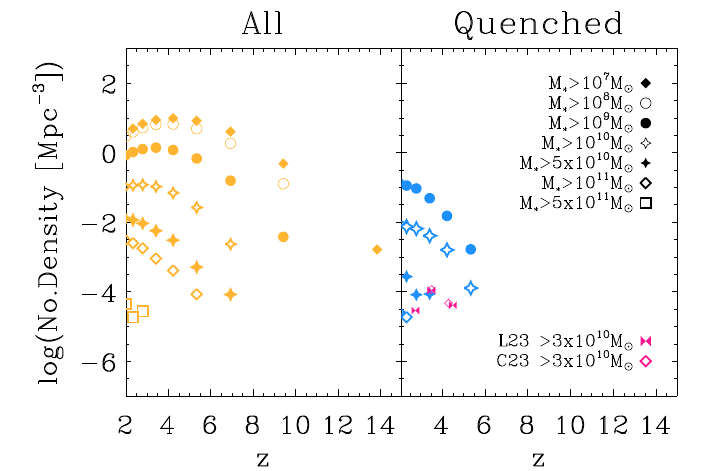}
    \caption{
    Number density of simulated galaxies above a given mass cut from $z=2$ to $z=15$. Different symbols mark the different lower mass limits applied, as indicated in the legend. \textit{Left:} All galaxies, regardless of their star formation properties. \textit{Right:} Only galaxies that are quenched.    Quenched galaxies exclude the highest mass bin, as none of the highest mass galaxies are yet quenched. They also do not include the two lowest mass cuts, as below $M_*<10^9M_\odot$ we cannot reasonably distinguish between quenched and non-quenched galaxies due to resolution limitations. Observations from \citet{long:2023} and \citet{carnall:2023} are included as pink bow-ties and diamonds, respectively.
    }
  {\label{fig:galdens}}
\end{center}
\end{figure}

Based on the observed stellar masses of the quiescent galaxies \citep{nanayakkara:2022,carnall:2023}, we require for this study that $M_*\geq3\times10^{10}M_\odot$, which equates to around $11000$~stellar particles per galaxy and results in sufficient resolution to study their properties in detail. This cut leaves us with $1309$~galaxies at $z=3.42$, the snapshot that most closely resembles the redshift at which quenched galaxies are currently observed, of which $1217$~are centrals. It is equivalent to the cut of level II from the companion study to this work by RK23.
\begin{figure*}
  \begin{center}
    \includegraphics[width=.9\textwidth]{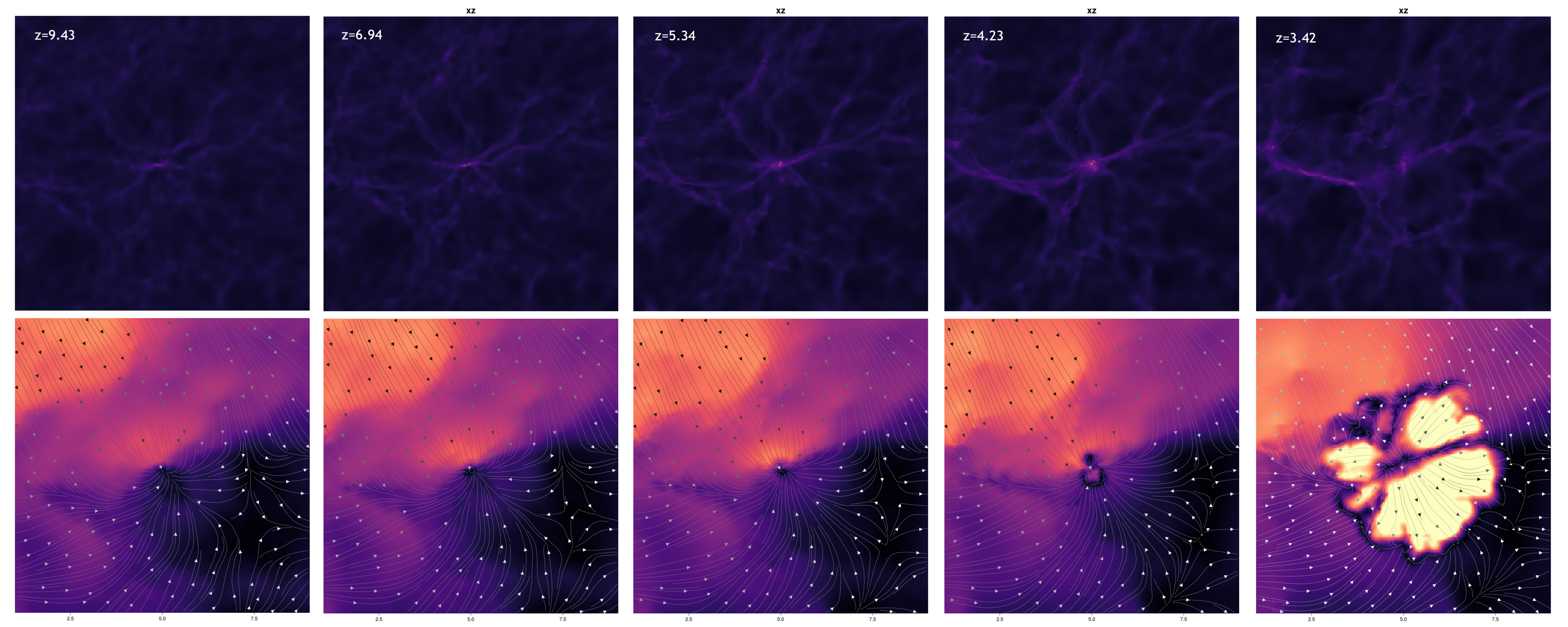}
    \caption{Projected gas density (\textit{top row}) and velocity (\textit{bottom row}) of galaxy~$ID16396$ at redshifts from $z=9.43$ down to $z=3.42$ (\textit{columns, left to right}), the moment when it is identified as quenched. 
    }
  {\label{fig:map}}
\end{center}
\end{figure*}

We determine a galaxy as 'quenched' if the current star-formation rate is equal to zero. At $z=3.42$, this results in $36$~quenched galaxies ($28$~centrals and $8$~satellites). We recover an identical sample when using the definition by \citet{franx:2008} based on specific star formation rate ($\mathrm{sSFR}=\mathrm{SFR} / M_*$). Quenched galaxies are then defined by $\mathrm{sSFR} < 0.3\times t_\mathrm{Hub}$. This alternative cut has been used in prior work for example by \citet{remus:2023} to predict the fraction of quiescent galaxies for different halo masses up to high redshifts comparable to those in this work, but also to study the quiescent population of galaxies at $z\approx2.7$ in comparison to observations \citep{lustig:2023}. Given the identical results, throughout this work we will stick to the strongest possible criterion for quenching, namely zero star formation.

Fig.~\ref{fig:galdens} shows the resulting number densities for all galaxies (left panel) and quenched galaxies only (right panel), for different redshifts from $z=15$ to $z=2$, for different stellar mass cuts from $M_*>10^7M_\odot$ (filled diamonds) to $M_*>5\times10^{11}M_\odot$ (open squares). As expected, small galaxies assemble the earliest, while the really massive galaxies with $M_*>5\times10^{11}M_\odot$ only appear at around $z=3-4$. This global behavior agrees well with observations. For the quenched fractions, we do not consider galaxies below $M_*<10^{9}M_\odot$, as for those galaxies quenching might be arbitrary due to resolution issues (where they would have less than $1000$~baryonic particles).
Nevertheless, it is very clear that quenched galaxies of a given mass appear later than their non-quenched counterparts of the same mass, clearly showing that the quenching process requires a certain amount of time to act on our galaxies.

As can also be seen from the right panel of Fig.~\ref{fig:galdens}, the most massive galaxies at a given redshift are not among the quenched ones, as there are no quenched galaxies in our sample with masses above $M_*>5\times10^{11}M_\odot$, and those with $M_*>10^{11}M_\odot$ barely appear around $z=2$. However, when comparing the quenched fractions from the simulations to observations by \citet{long:2023} and \citet{carnall:2023}, we find good agreement, indicating that the timescales of quenching in our simulation act on comparable scales to the real one.

In the following, we will now restrict our analysis to those $36$~galaxies that are found to be quenched at $z=3.42$, $28$~of which are centrals at that redshift, and $22$~of those stay central galaxies until $z=2$. Whenever galaxy properties are traced through time, these $22$~galaxies are considered unless stated otherwise. They represent the core sample of this study, with a few special cases introduced specifically. All galaxies that are not quenched at $z=3.42$ and remain central galaxies until $z=2$ are used as non-quenched control sample throughout.
In particular, the sample of galaxies at $z=3.42$ consists of $1217$~galaxies which are centrals, with $1189$~of them that are not quenched. Throughout this study we refer to the $28$~quenched central galaxies at $z=3.42$ as 'quenched', and the remaining centrals as 'non-quenched', and note that this does not refer to their current state at all redshifts but rather to what they eventually are at $z=3.42$.

\section{How Do Galaxies at High Redshifts quench?}\label{sec:quench}
Having seen that the global population of massive galaxies is well reproducing the observations at high redshifts, we turn here to the central question of how massive quiescents form and quench. A first indication can be gained from Fig.~\ref{fig:map}, which shows the projected gas density (top) and velocity (bottom) of the surroundings of one of our quenched galaxies, galaxy~$ID16369$. It is one of the most massive central galaxies which quenches by $z=3.42$, reaching a stellar mass of $M_*\approx5.6\times10^{10}M_\odot$. 

Going from left to right we find that galaxy~$ID16369$ collapsed as one of the earliest structures in its surroundings, present as an overdensity already at $z=9.43$ (leftmost column). Consequently, the surrounding gas is primarily flowing toward it. Interestingly, the gas inflow is largely isotropic with the exception of the bottom right, where there is a noticeable underdensity. As discussed further in Sec.~\ref{sec:env}, we find that lying in or near an underdense region is actually not unusual for the quenched galaxies, but rather a requirement.

This accretion continues until $z=4.23$, which is the last moment prior to the galaxy quenching. We note that a few minor filaments have formed by this point through which gas is flowing, while to the left another overdensity is beginning to collapse. A small outflow can already be seen to emerge from the galaxy in the flow field, although no impact on the gas density is yet visible.

It is then in between $z=4.23$ and $z=3.42$ where the galaxy quenches. This process is rapid and violent, resulting in a massive expulsion of gas away from the galaxy (bottom right panel). We find this expulsion to be sufficiently powerful to disrupt the thin connecting filaments in the gas (top right panel), which become much more diffuse. Furthermore, the ejected gas can be driven out particularly far into the aforementioned underdensity located in the lower right. 
Though this is just an example of one of the quenched galaxies, it illustrates the process found for all of them, as we will quantify more in the following.

\subsection{Inside-Out Quenching}\label{subsec:radial}
\begin{figure*}
  \begin{center}
    \includegraphics[width=.99\textwidth]{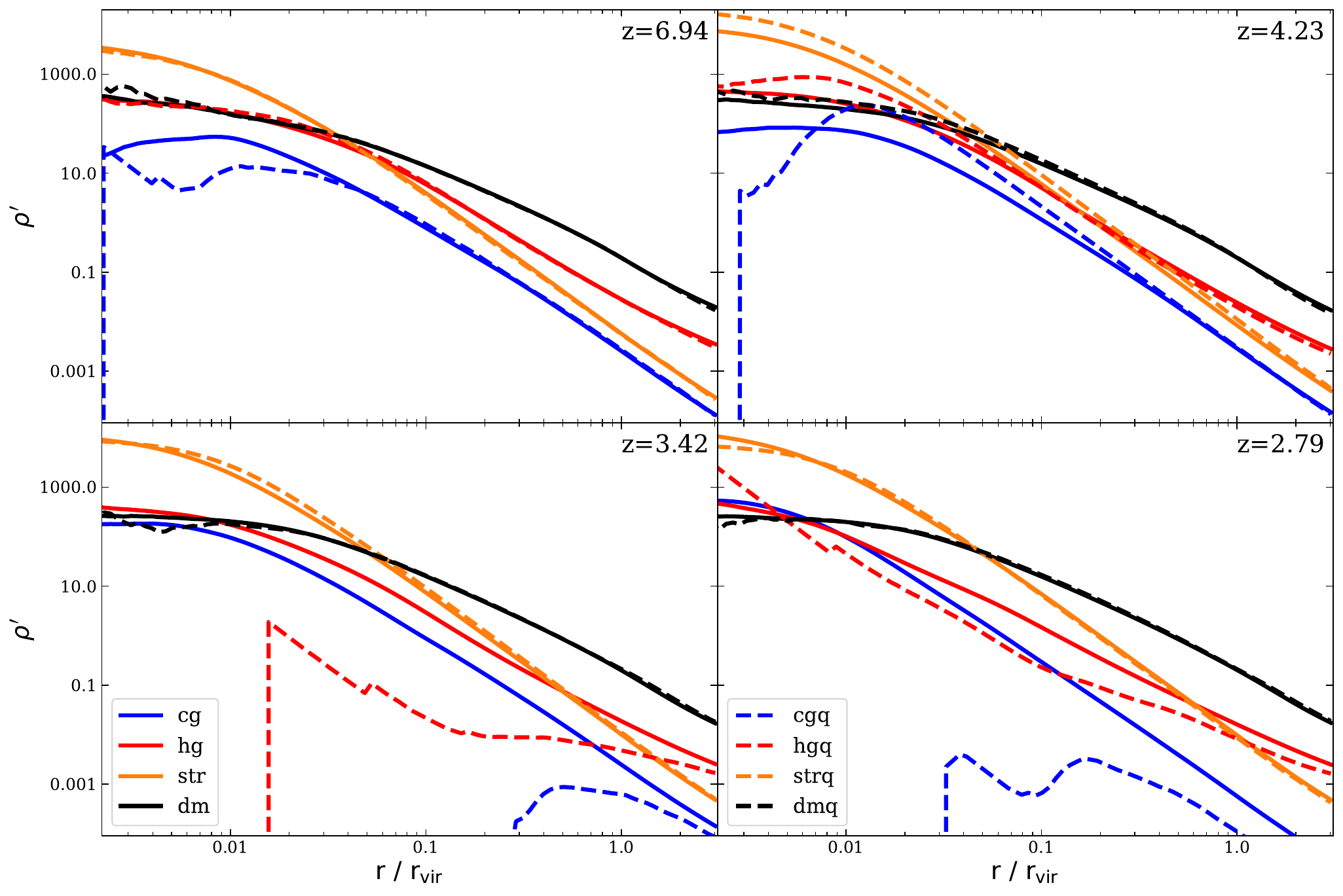}
    \caption{Stacked and normalized density profiles of the cold gas ('cg', blue), hot gas ('hg', red), stellar ('str', orange) and dark matter ('dm', black) components within the halos of quenched (dashed lines) and non-quenched centrals (solid lines) for four different redshifts as indicated in the top right. 
    }
  {\label{fig:radial_rho}}
\end{center}
\end{figure*}
To understand the process of quenching in more detail, all $1217$~centrals at $z=3.42$ are traced forward and backward in time, keeping only those which at the given redshift are centrals. This results in $1213$,~$1201$,~$1217$ and $1078$~centrals at $z=6.94$, $z=4.23$, $z=3.42$ and $z=2.79$, respectively. For these galaxies, we first want to understand whether or not there is a difference in the mass distribution present already at high redshifts before quenching.

Therefore, we determine for each galaxy at a given redshift the density profiles for the stars, dark matter, and cold ($T\leq10^4K$) and hot ($T>10^4K$) gas. To remove variations in absolute density or radius, we normalize every profile to the galaxy's local density (baryonic and dark matter) at the virial radius as $\rho'_i\equiv\rho_i/\rho_\mathrm{tot}(r_\mathrm{vir})$. This allows us to quantify the \textit{shape} of the profiles of different components of the quenched versus non-quenched galaxies. 

Fig.~\ref{fig:radial_rho} shows the evolution of these normalized profiles at all four redshifts. At a high redshift of $z=6.94$, we find little difference in the shapes of the dark matter (black), stars (orange) or hot gas (red) between the non-quenched (solid lines) and quenched galaxies (dashed lines). There is a slight dip in the density of the cold gas (blue) in the quenched progenitors at around $r/r_\mathrm{vir}\approx0.01$, however, this difference is small. 

At a redshift of $z=4.23$, the moment prior to quenching, we find noticeable differences between the two samples (top right of Fig.~\ref{fig:radial_rho}). The quenched galaxies are going through a massive starburst relative to the non-quenched sample. This results in an increased concentration of the central stellar component for $r<0.1\times r_\mathrm{vir}$ where the orange dashed line lies above the orange solid line, corresponding to a dip in the cold gas density (seen at $r/r_\mathrm{vir}<0.01$, around $0.6\mathrm{kpc}$). 

Farther out at $r/r_\mathrm{vir}>0.01$ there is a ``shockwave'' of ejected cold and hot gas characterized by a noticeable density peak. It is interesting to note that the hot gas peak is slightly farther in radially, which may be indicative of star formation feedback (which is coupled to the cold gas) triggering first before the AGN feedback begins driving both gas phases out. This also agrees well with the small-scale outflow visible in Fig.~\ref{fig:map} at $z=4.23$ for our example galaxy~$ID16369$.

Post quenching, i.e. at $z=3.42$ (bottom left of Fig.~\ref{fig:radial_rho}), the quenched galaxies have ejected all hot and cold gas beyond $0.01$ and $0.3\times r_\mathrm{vir}$, respectively. Even where gas is present, the relative density of the gas is reduced by around an order of magnitude relative to the non-quenched sample. Clearly, the process of quenching is accompanied by a massive feedback burst from both the stellar and the AGN components of the galaxy, which results in quenched centrals ejecting their gas out to large radii (see Sec.~\ref{subsec:stellar}).

Curiously, there is a minor underdensity in the most central dark matter component of the quenched galaxies, and the stellar core reaches farther out. This could be an indication for feedback-induced halo expansion as discussed for example by \citet{dutton:2016}, \citet{peirani:2017}, and \citet{harris:2020}.
Note that going from $z=4.23$ to $z=3.42$ has subtly changed which regions of the halo are being probed. At $z=4.23$ both the halos which host quenched as well as non-quenched centrals have a median virial radius of $\bar{r}_\mathrm{vir}\approx 55\mathrm{kpc}$. In contrast, at $z=3.42$ the halos of quenched centrals have $\bar{r}^\mathrm{q}_\mathrm{vir}=69\mathrm{kpc}$, while non-quenched have $\bar{r}^\mathrm{nq}_\mathrm{vir}=79\mathrm{kpc}$.

Following this violent quenching process, gas starts flowing back into the quenched central as shown in the bottom right of Fig.~\ref{fig:radial_rho}. The hot gas in particular forms a strongly peaked profile, while the cold gas reaches towards the centers of the galaxies again but remains noticeably underdense. The dark matter component has settled to largely match the shape within the non-quenched centrals. Interestingly, the stellar component has become slightly more shallow towards the center, however, that imprint is small.
\begin{figure}
  \begin{center}
    \includegraphics[width=.9\columnwidth]{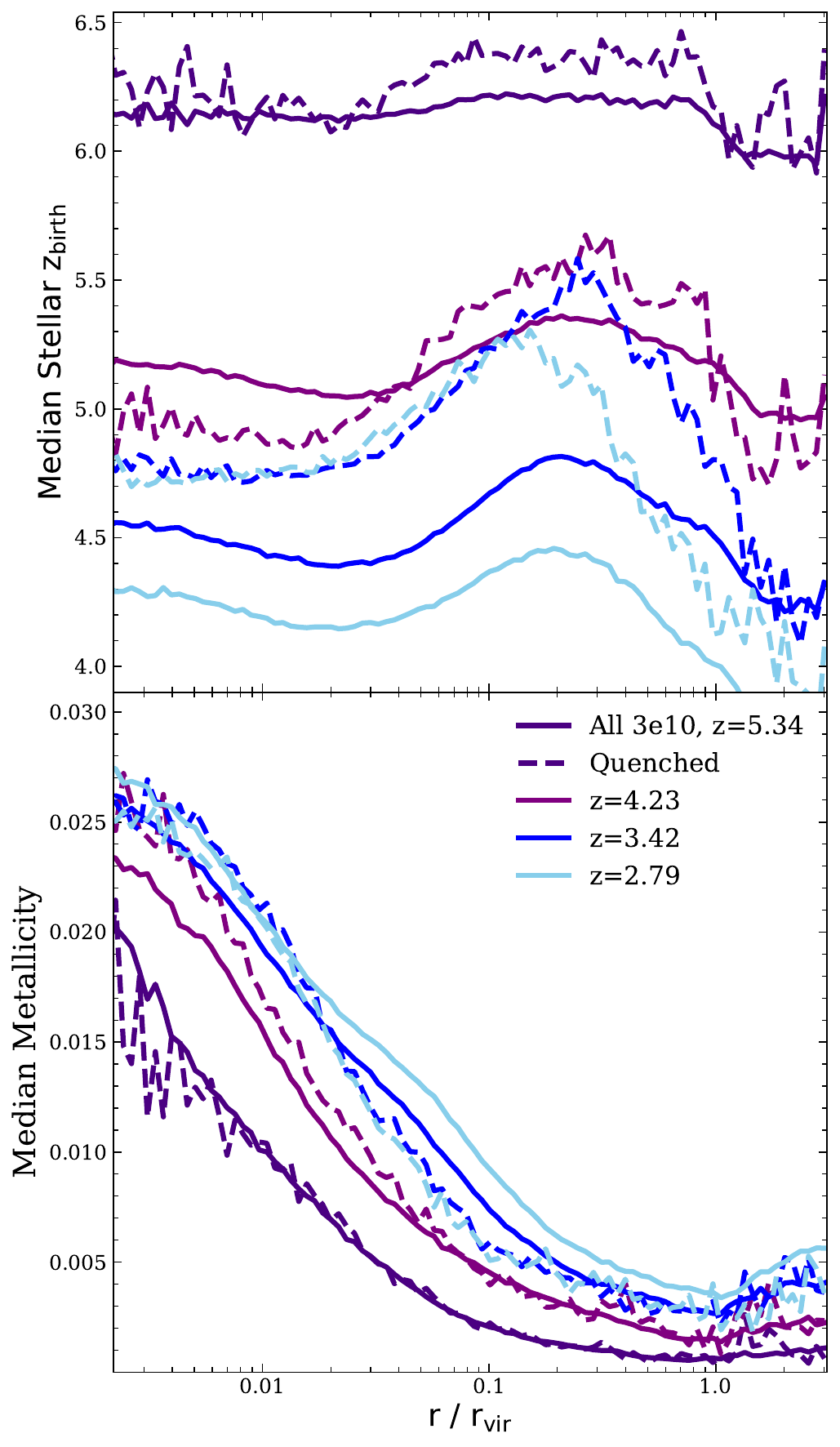}
    \caption{Stacked quenched (dashed) versus non-quenched (solid) radial profiles of median stellar $z_\mathrm{birth}$ (\textit{top}) and metallicity (\textit{bottom}) for different redshifts (colors) as given in the legend. 
    }
  {\label{fig:age_metal_radial}}
\end{center}
\end{figure}
To summarize, we find that the overall amount of stars and their radial distribution show no significant differences between quenched and non-quenched galaxies, indicating that, despite the star formation process being more rapid for the quenched galaxies, the overall star formation takes place with a similar resulting mass distribution, just on different timescales. Nevertheless, the non-quenched galaxies continue forming stars, thereby continuously varying their radial profiles, while the quenched galaxies will remain as such so long as they do not rejuvenate (see RK23 for this scenario).

The observed temporal difference in the profiles should, however, be imprinted in the properties of the stars, as a short starburst has different properties with respect to the metal content than a slow, but long-ongoing star formation. To see this impact, Fig.~\ref{fig:age_metal_radial} shows the stacked profiles of the median stellar birth redshift $z_\mathrm{birth}$ (upper panel) and median metallicity (lower panel) as a function of radius. Metallicity is defined here as the summed mass of all implemented metals \citep[see][]{dolag:2017} divided by the hydrogen mass. 

At $z=5.34$ (dark purple lines), the quenched and non-quenched samples show largely similar trends, with flat stellar age profiles and smoothly increasing metallicities towards the center. Going to $z=4.23$ (light purple), the quenched galaxies (dashed lines) exhibit a younger and more metal rich population in the centers, with a somewhat older stellar halo compared to the non-quenched sample. We interpret this as the result of intense funneling of gas towards the centers (see Sec.~\ref{sec:env}), which causes a starburst that reduces the stellar ages and also, through feedback from supernovae Type II, metal enriches the gas and thus also the metallicity of the newly formed stars. This is in agreement with the radial stellar density profiles of the quenched galaxies being denser than for the non-quenched sample. 

After the burst and subsequent quenching, at $z=3.42$ (blue) the central core $r/r_\mathrm{vir}<0.01$ of the quenched galaxies remain comparatively metal enhanced. Due to the cessation of star formation in the quenched galaxies, however, the metallicity in the galaxies' outskirts is now comparatively lower, and the overall ages of the non-quenched sample are younger, showing the opposite trend to what was seen prior to quenching. It is only in the regions beyond the virial radius where the stellar population remains comparable in both age and metallicity for the quenched and non-quenched galaxies. 

As most of the quenched galaxies remain quenched for some time (see the discussion by RK23), their central core has practically identical ages and metallicity at $z=2.79$ (light blue) compared to $z=3.42$. Intriguingly, for $r>0.1\times r_\mathrm{vir}$ there is a build up of younger metal-poor stars, lowering the median $z_\mathrm{birth}$. We interpret this as the initial build-up of a stellar halo, and when tracing down to $z=2$ we find that the quenched galaxies exhibit a relatively flat age gradient before becoming negative when transitioning into the younger halo. 

\begin{figure*}
  \begin{center}
    \includegraphics[width=.9\textwidth]{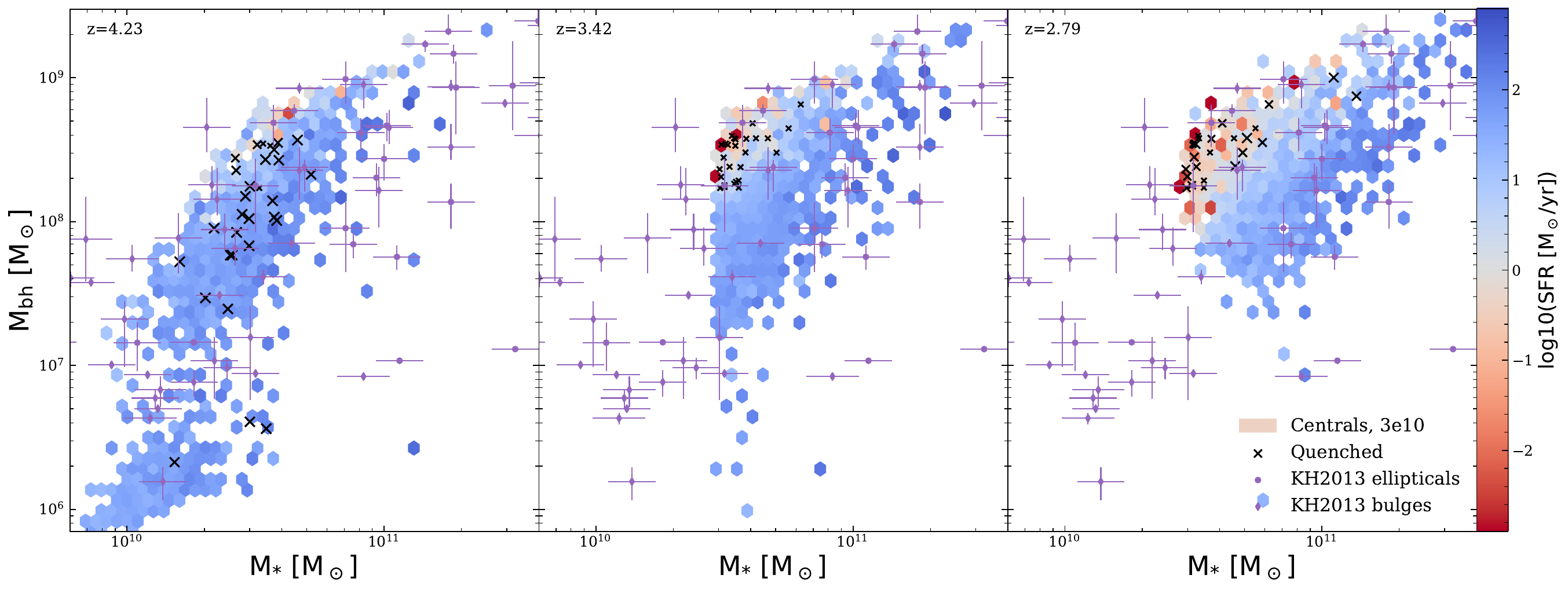}
    \caption{Black hole mass versus stellar mass of the central galaxies in our sample at a redshift of $z=4.23$ (\textit{left}), $3.42$ (\textit{middle}) and $2.79$ (\textit{right}), colored by star formation rate. Quenched galaxies are plotted as black crosses with the size increasing with star formation rate. Observed galaxies at $z=0$ by \citet{kh13} are shown in purple for comparison.
    }
  {\label{fig:agntest}}
\end{center}
\end{figure*}

Effectively, the stellar ages and metallicities of the quenched galaxies are frozen in at the moment of quenching, and as long as the quenched galaxies avoid rejuvenation or otherwise dry accretion, this imprint of their formation scenario should be kept and still visible at present-day. Such galaxies would be forming what we at present day call the relic galaxies \citep{spiniello:2021,spiniello:2023}, but could in case of rejuvenation or continuing merger processes simply be part of the old stellar population of a present-day galaxy.

\subsection{Combined Stellar and AGN Feedback}\label{subsec:stellar}

As discussed in the introduction, the actual quenching mechanisms at high redshifts are still under debate. While IllustrisTNG predicts quenching to result solely from AGN feedback \citep{hartley:2023,kurinchi:2023}, models and observations are not as set on the AGN scenario as the sole reason for quenching; they rather find a co-evolution between high star formation rates in short bursts and evidence of AGN activity \citep{tacchella:2016,tozzi:2023,dsilva:2023,fan:2023}.
Furthermore, simulations from NIHAO have shown that supernova feedback is also able to generate strong outflows that even reach beyond the virial radius \citep{tollet:2019}. Given the strong nature of the observed starbursts, the feedback from supernovae expected at high redshifts is most likely non negligible.
This raises the question whether the feedback from the central black hole is sufficient by itself to explain the quenching of galaxies at high redshifts, or if other components also play a crucial role. 

\begin{figure*}
  \begin{center}
    \includegraphics[width=.88\textwidth]{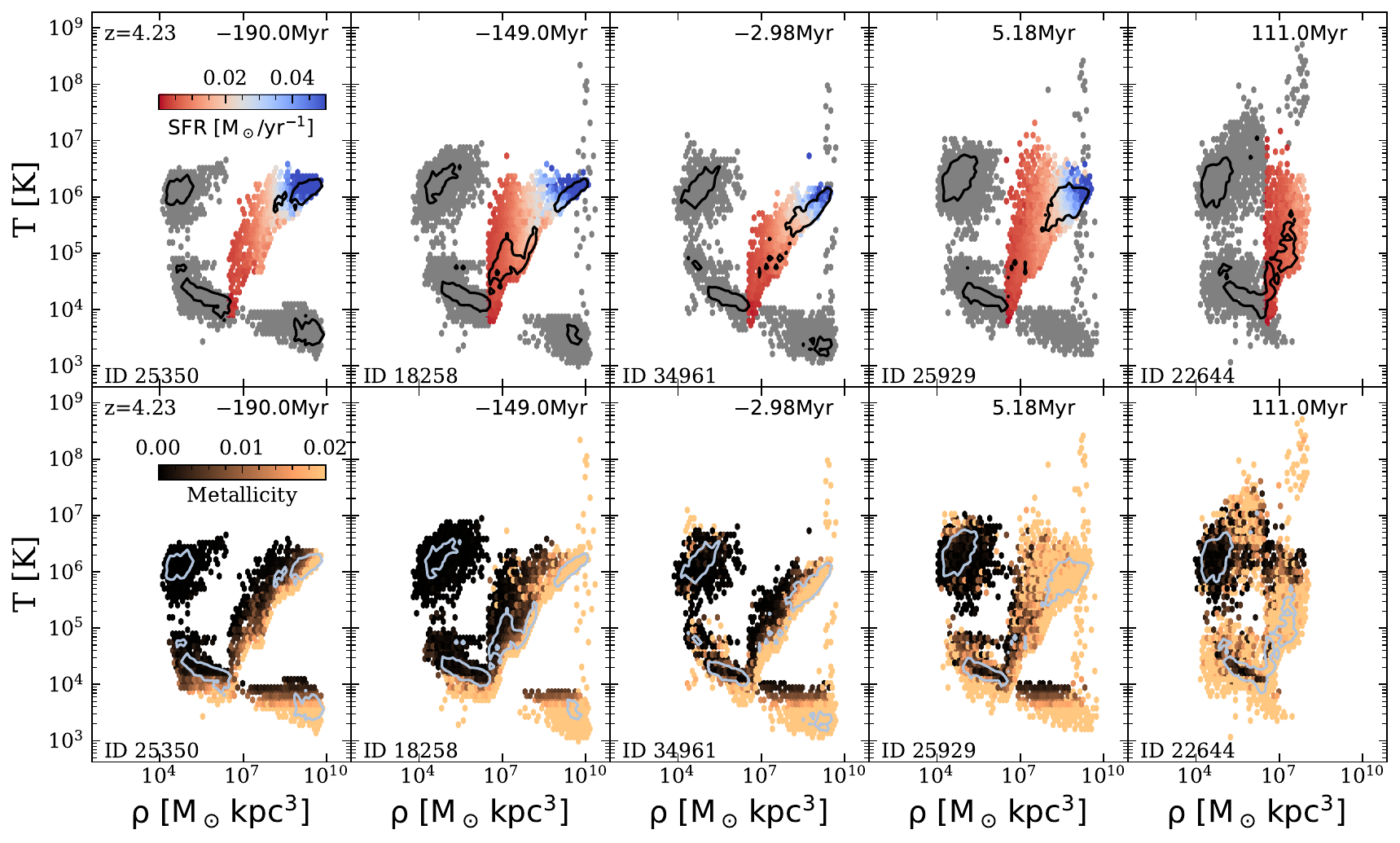}
    \includegraphics[width=.88\textwidth]{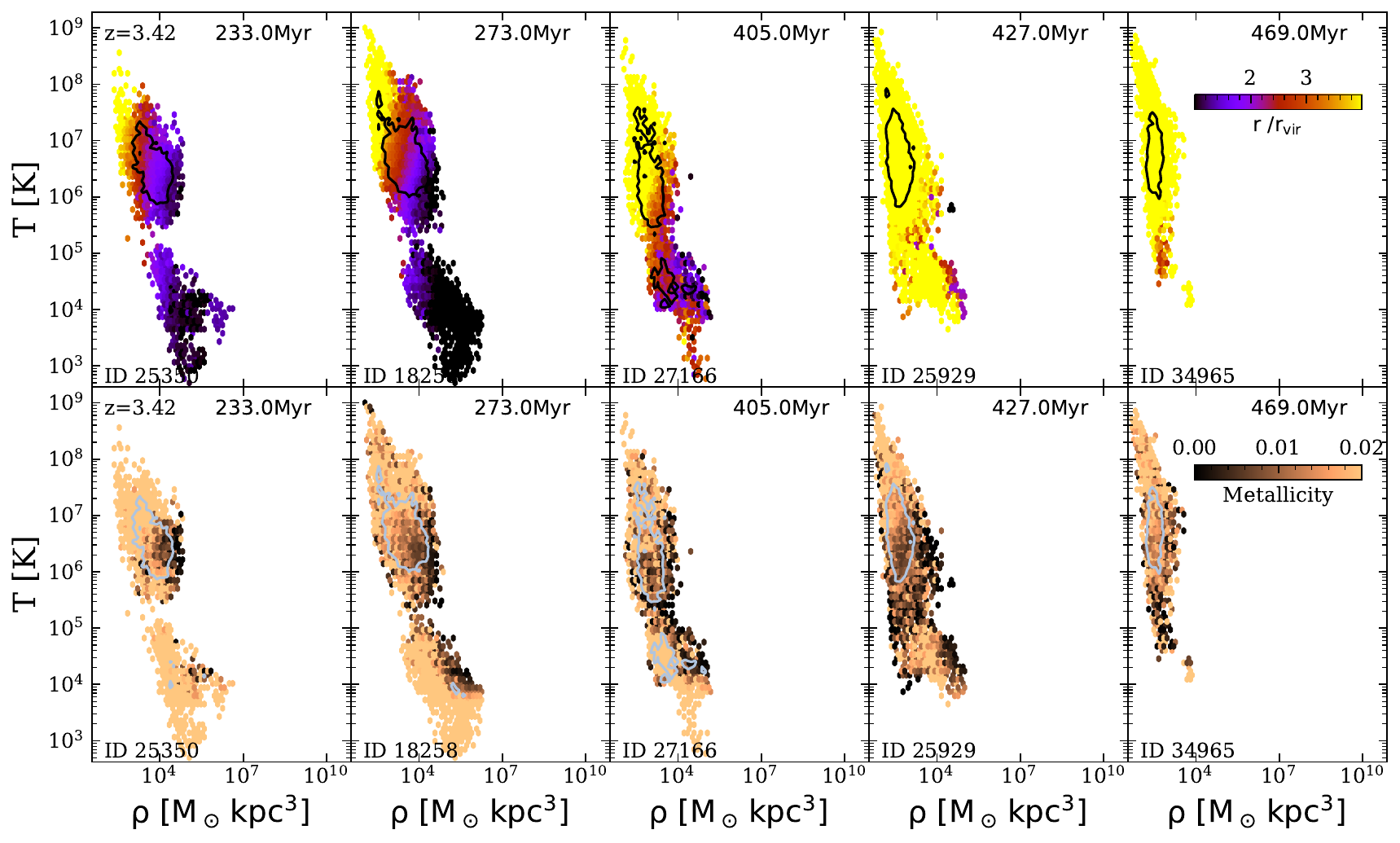}
    \caption{Phase space evolution of quenched galaxies' gas prior to (\textit{top}, $z=4.23$) and post quenching (\textit{bottom}, $z=3.42$). Galaxies shown from left to right are sorted based on their $t_{90}$ (given in the top left corner), with negative values implying that they have yet to form $90\%$ of their final stellar mass present at $z=3.42$. Contours enclose half the gas. IDs of the galaxies are given in the bottom left.
    }
  {\label{fig:phase}}
\end{center}
\end{figure*}

To answer this question, we first consider the well-known black-hole-to-stellar-mass relation at redshifts from $z=4.23$ (pre-quenching) to $z=2.79$ (post-quenching) in Fig.~\ref{fig:agntest}. We plot here only central galaxies to avoid including additional effects from the environment, and only those which have a stellar mass of $M_*>3\times 10^{10}M_\odot$ at $z=3.42$ to be comparable with the observed quenched galaxies.

We find that the simulated high redshift galaxies overall follow the relation found in observations at $z=0$ \citep{kh13}. This is in agreement with observations by \citet{ding:2022} who found a black hole in a quasar at $z\approx6$ to already reside on the black-hole-to-stellar-mass relation observed at present day. However, recently \citet{pacucci:2023} and \citet{harikane:2023AGN} presented evidence from JWST observations at z=4-7 for this relation to be shifted towards higher black hole masses for a given stellar mass, albeit this is still under debate. 
Note that, as our model seeds central black holes at masses below the relation (which can be seen for some of the galaxies at $z=4.23$, which are gathering in a cloud in the lower left of Fig.~\ref{fig:agntest}), this implies that the SMBHs first rapidly grow until they reach the relation, after which they evolve according to it. 

When coloring by star formation rate we find that galaxies hosting more massive black holes at equal stellar masses generally have a lower star formation rate (redder color). However, the correlation does not hold for all galaxies, with numerous galaxies lying at the highest black hole masses and yet still exhibiting significant star formation. 

The quenched galaxies (black crosses) host particularly massive SMBH by the time they are identified as quenched ($z=3.42$), but they do not represent significant outliers. It should be noted that AGN feedback scales not only with the absolute mass but also with the recent mass accretion \citep{steinborn:2015}. However, also in terms of SMBH mass growth there are $197$ centrals which grow more significantly than the median growth of quenched galaxies (both in absolute mass as well as mass ratio between $z=4.23$ to $z=3.42$).
This means that the pure growth of a massive central black hole by itself is insufficient to cause a complete stop of star formation. This is in agreement with the observation of active AGN in galaxies with ongoing star formation for example by \citet{tozzi:2023}.

Thus, if the AGN feedback alone is not the sole reason for quenching, the question remains what else contributes. Therefore, we turn to analyze what happens to the gas within a galaxy during the process of quenching.
Due to the limited time resolution of the stored simulation outputs we cannot follow a single galaxy through the process from start to finish, despite the original simulation of course performing all this in detail. We can, however, quantify how far along the quenching of a given galaxy is by using $t_\mathrm{90}$, the time since a galaxy made $90\%$~of its final stellar mass at $z=3.42$. A negative $t_\mathrm{90}$ implies that the galaxy has yet to make $90\%$.

The top set of panels of Fig.~\ref{fig:phase} shows the phase space diagram of the gas within the virial radius at $z=4.23$ for five centrals which will quench by $z=3.42$. Going from left to right they are sorted from lowest to highest $t_\mathrm{90}$, and the gas particles are colored by star formation rate (upper) and metallicity (lower). 
We find that around $200\,$Myr before reaching $t_\mathrm{90}$ the galaxy is undergoing a massive starburst, with a prominent star forming branch. This results in a noticeable cloud of particularly cold and dense gas particles, so-called wind particles. They are the result of gas particles being subjected to stellar feedback and winds.

As the starburst progresses (moving toward the right panels) the star-forming branch begins to be disrupted, with gas particles being heated due to the feedback processes. In particular we note multiple metal rich gas particles which start being found at lower densities of $\rho\approx10^5 M_\odot \mathrm{kpc}^{-3}$. At $t_{90}=-2.98\,\mathrm{Myr}$ this ejected metal rich gas appears also in the region of phase space occupied by the IGM (thin and hot). The fact that this gas is metal enriched indicates that it has been subject to stellar feedback, either through SN or ABG stellar winds. 

At $t_{90}=5\,\mathrm{Myr}$ the star forming branch is now significantly disrupted, until finally at $t_{90}\approx 100\,\mathrm{Myr}$ all gas has been removed from the most dense central regions, either by being used up for star formation or by being ejected. We note also the presence of a cloud of hot and metal rich gas which is in the process of being ejected. 

Tracing the gas which was present within the galaxy forward to times post quenching (bottom set of panels), we find two main properties: the ejected gas is noticeably metal enhanced (bottom row) and reaches out to significant radii in excess of $r>3\times r_\mathrm{vir}$ (top row). At $t_{90}>400\,\mathrm{Myr}$ all gas which was present pre-quenching has been fully removed from the galaxy. 

This again is in good agreement with what we found for our example halo in Fig.~\ref{fig:map}, where we saw the gas being driven outwards after quenching. However, the fact that the gas that is being ejected is already metal enriched clearly shows that the starburst precedes the ejection of the gas through feedback during quenching, clearly indicating that the stellar feedback from the supernovae Type II adds to the energy budget of the quenching process, and that a massive starburst is required preceding the quenching.

\subsection{Massive Quiescents Quench Quickly}\label{subsec:rapid}

In the previous section we have shown how quenched galaxies exhibit a starburst preceding the enrichment and ejection of all gas which ultimately results in a stop to all star formation (i.e., quenching). We turn here to the question on which timescales this occurs and whether there is any signature left behind in the quenched galaxy which could be observed. 

\begin{figure*}
  \begin{center}
    \includegraphics[width=.9\textwidth]{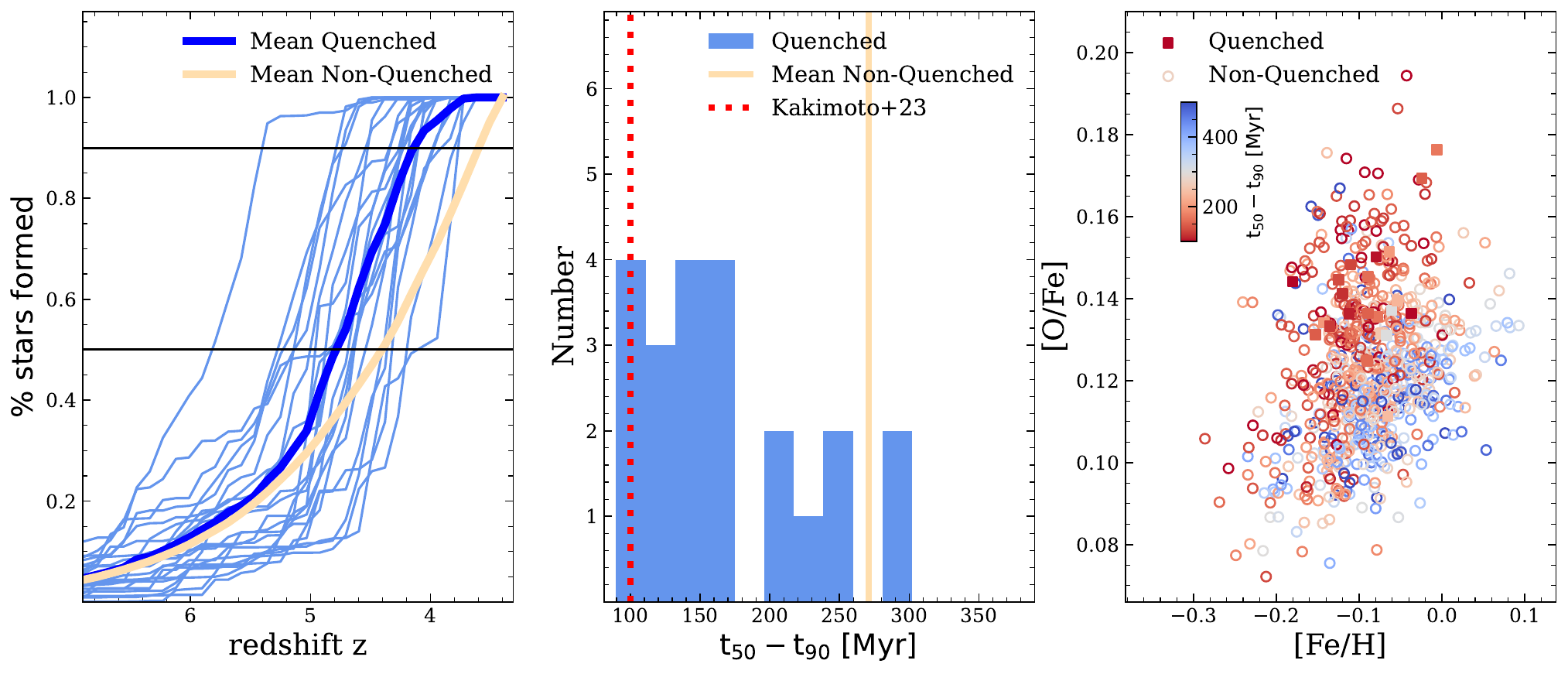}
    \caption{\textit{Left:} Star formation history of quenched centrals (thin blue lines) as well as their mean (thick blue line) and the mean of non-quenched centrals (thick sand line). \textit{Center:} Distribution of the quenching times $t_\mathrm{50}-t_\mathrm{90}$ for quenched centrals (blue bars), as well as the mean time for non-quenched (sand color line) and the quenching timescale given by \citet{kakimoto23} (red dotted line). \textit{Right:} Alpha enhancement given by [O/Fe] versus metallicity [Fe/H] for all centrals colored by $t_\mathrm{50}-t_\mathrm{90}$, with redder colors indicating a more rapid burst. Quenched centrals are plotted as filled squares.
    }
  {\label{fig:shist}}
\end{center}
\end{figure*}

The left panel of Fig.~\ref{fig:shist} shows star formation histories (SFH) for the quenched centrals (thin blue lines). There emerges a characteristic 'S'-shape, with an initial period of low star formation, followed by a burst producing around $80\%$ of the final stellar mass and subsequent rapid quenching (flattening) of the SFH. The difference to the non-quenched galaxies can be seen clearly by comparing the mean of the quenched centrals (thick blue line) to the mean of the non-quenched centrals (thick sand line). 

To characterize how rapidly this starburst occurs, we consider for a galaxy the elapsed time between forming $50\%$ and $90\%$ (horizontal black lines in the left panel) of its final stellar mass at $z=3.42$, $t_\mathrm{50}-t_\mathrm{90}$.
The resulting times are shown in the middle panel of Fig.~\ref{fig:shist}. We find that most of the quenched galaxies have $t_\mathrm{50}-t_\mathrm{90}<170\,\mathrm{Myr}$. Four quenched galaxies of the sample even reach times around~$100\,\mathrm{Myr}$, comparable to the quenching timescale observed by \citet{kakimoto23} (red dotted line), but also by \citet{carnall:2023onequenchie} (approx. $200\,$Myr). This differs clearly from the non-quenched galaxies, for which the mean $t_\mathrm{50}-t_\mathrm{90}$ is nearly at $300\,$Myr. While this still is a rather short time span at present day, it actually is a large difference at such extreme high redshifts, and it is expected as the non-quenched galaxies at $z=3.42$ studied here are also fairly massive and had to form stars in a rather short period of time. Interestingly, the starburst timescales found in our simulation are similar or even slightly shorter than those found for the quenched galaxies from the FLARES simulations at high redshifts by \citet{lovell:2023b}.

Alternatively, we can determine the time between the last formed stellar particle and the highest peak in SFR, termed $t_q$. We bin the stellar formation times in steps of equal time and define $t_q$ then as the time between the bin with the highest formed mass and the last bin containing any mass. Note that this definition is dependent on the chosen size of bin, so we test for bins of time $10\,\mathrm{Myr}$ to $100\,\mathrm{Myr}$. 

We find that $t_q$ ranges from $30\,\mathrm{Myr}$ to $560\,\mathrm{Myr}$. The mean lies at $t_q\approx 183\,\mathrm{Myr}$, and is largely stable for different sizes of bin (lying between $160\,\mathrm{Myr}$ and $200\,\mathrm{Myr}$). Due to the more sporadic shape of individual bursts in SFR, we choose here to focus on the time characterizing the more smooth build up of stellar mass, $t_\mathrm{50}-t_\mathrm{90}$, instead.

Returning then to $t_\mathrm{50}-t_\mathrm{90}$, we find that the quenched galaxies experience a generally more rapid burst. The rapidity of this process should result in a stellar population which is enriched primarily through the faster SNII compared to SNIa. This enhances alpha elements such as oxygen faster than iron. 

We define [X/Y]$=\log_{10}(\mathrm{M_X/M_Y}) - \log_{10}(\mathrm{M_X/M_Y})_\odot$, with solar values given by \citet{wiersma:2009}. The right panel of Fig.~\ref{fig:shist} shows the locations of the stellar populations for all central galaxies in the [O/Fe]-[Fe/H] plane at $z=3.42$, colored by the rapidity of the star formation given by $t_\mathrm{50}-t_\mathrm{90}$. 
As is expected, we find that galaxies which underwent more rapid star formation (shorter $t_\mathrm{50}-t_\mathrm{90}$) lie at higher [O/Fe] compared to those of equal [Fe/H] which exhibit a more prolonged, continual star formation. The quenched galaxies (filled squares) reside in the area of highest alpha enhancement with comparable [Fe/H] to the non-quenched centrals. Consequently, the observed relation between alpha enhancement and central surface mass density in early type galaxies (ETGs) at lower redshifts \citep{barone:2018} may well in part result from galaxies quenched via the mechanism described here, producing compact and strongly alpha enhanced galaxies. We conclude that using the [O/Fe]-[Fe/H] plane as a diagnostic for the recent star formation history is feasible also at redshifts $z>3$.  

\begin{figure}
  \begin{center}
    \includegraphics[width=.9\columnwidth]{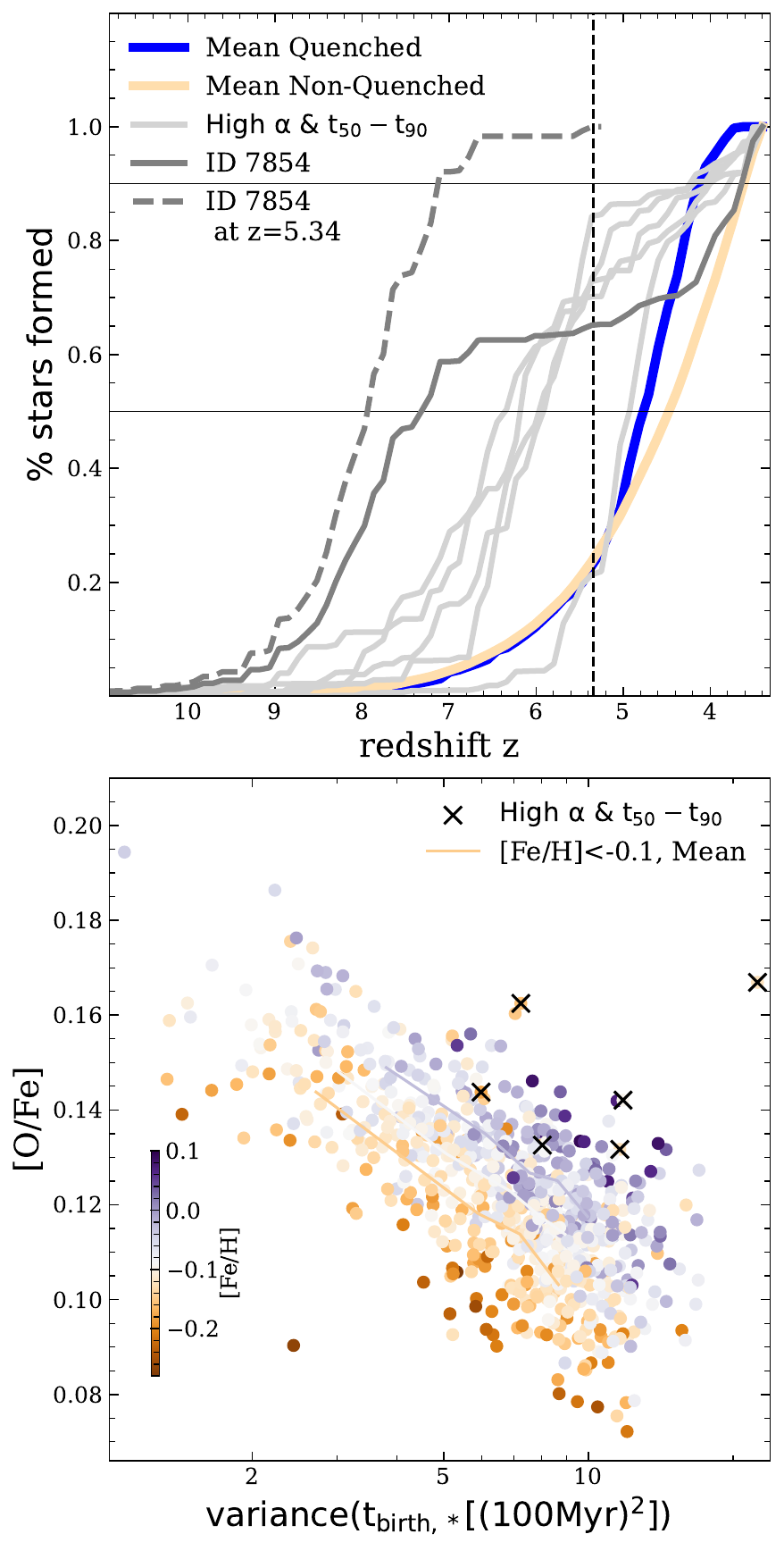}
    \caption{\textit{Top:} Star formation histories for the six outliers (gray), with one particular case highlighted in a darker gray. For this galaxy the SFH is shown also at $z=5.34$ (gray dashed). Mean SFH for the quenched (thick blue) and non-quenched (thick sand) are shown for comparison. \textit{Bottom:} Stellar alpha enhancement [O/Fe] versus the variance in birth times of stellar particles, colored by metallicity [Fe/H]. Mean trends for low, intermediate and high [Fe/H] are given as colored lines (orange, silver, purple). The six outliers are indicated by black crosses. 
    }
  {\label{fig:early}}
\end{center}
\end{figure}
However, there are a few notable outliers with enhanced [O/Fe] yet particularly large $t_\mathrm{50}-t_\mathrm{90}$. We identify six such galaxies which have [O/Fe]$>0.13$, [Fe/H]$<-0.09$ and $t_\mathrm{50}-t_\mathrm{90}>500\,\mathrm{Myr}$. 
The star formation histories of these six outliers down to $z=3.42$ are shown as solid gray lines in the top panel of Fig.~\ref{fig:early}, with the mean trends of quenched (blue) and non-quenched (sand) centrals for comparison. Five of the outliers show the same characteristic 'S'-shape of the quenched galaxies, but shifted to earlier times. They indeed undergo a rapid phase of star formation (hence the enhanced [O/Fe]) but stop just short of forming $90\%$ of their stars. Instead, they transition into a more gradual phase of star formation which results in a high $t_\mathrm{50}-t_\mathrm{90}$. This post starburst phase, however, is not extended enough to significantly enhance [Fe/H]. 

The last of the six galaxies, galaxy~$G7854$, is a particularly interesting case. Its SFH shows that it underwent \textit{two} separate star formation bursts, with a prolonged period of relative quiescence in between. In fact its first burst is so early that, were it observed at $z=5.5$, its formation redshift would be $z_\mathrm{form}\approx 8$ (dashed gray line), and it would appear both quiescent and massive ($M_*>1\times10^{10}M_\odot$), making it comparable to all but the earliest forming quiescent galaxy from \citet{nanayakkara:2022}, similar to the results for the quenched galaxy reported by \citet{glazebrook:2023}. 

It follows that the original assumption that enhanced [O/Fe] is the result of a phase of rapid star formation holds also for these outliers. It is through the peculiar shape of their SFH that they posses unusually large $t_\mathrm{50}-t_\mathrm{90}$. 
This can be seen further in the bottom panel of Fig.~\ref{fig:early}, where we show [O/Fe] as a function of the variance in stellar age. Low variance indicates that most of the stars were formed in a single star formation event and correlates with enhanced [O/Fe]. On the other hand, high variance indicates more prolonged star formation, and subsequently more iron enrichment through SNIa, which reduces [O/Fe]. The six outliers inhabit here the region of particularly high [O/Fe] and high age variance, with the most extreme outlier with the highest variance being galaxy~$G7854$. 

Furthermore, we find that the scatter in [O/Fe] at a given variance in stellar age is driven by different total metallicities [Fe/H]. Interestingly, higher [O/Fe] are found for galaxies with higher [Fe/H]. We interpret this as a result of the amount of self enrichment which occurs within a galaxy. As star formation at high redshift is primarily bursty, metal enrichment is driven by the SNII. Consequently, a galaxy which manages to enrich its star-forming gas more efficiently in [Fe/H] will be doing so mostly through SNII, thus driving [O/Fe] even higher. 

This imprint of the metal enrichment in the alpha elements should, if a galaxy is quenched at high redshifts and does not undergo either merging or rejuvenation, still be measurable in the galaxy if observed at present day. Thus, if relic galaxies like those found by INSPIRE \citep{spiniello:2021,spiniello:2023} and \citet{comeron:2023} really are the remnants of quenched galaxies at cosmic dawn, they should not only be rather compact and have extremely old stellar ages, they should also show strong alpha enhancements in their metal composition.

\subsection{Baryon Fraction and Kinematics of Quenching}\label{subsec:kin}
We have shown that the radial profiles of the stars and dark matter of quenched versus non-quenched galaxies at $z=5-2$ do not differ strongly in their overall behavior. However, it has been claimed that quenched galaxies reside in comparatively overmassive halos and could have enhanced baryon fractions. We do not find our quenched galaxies to live in particularly massive halos (see RK23 for more details), in fact the most massive halos at $z=3.42$ all host star forming central galaxies. This is vastly different from the results found for IllustrisTNG by \citet{hartley:2023} and \citet{kurinchi:2023}, where the quenched galaxies only reside in the most massive halos, as only in such halos the AGN has already switched to kinetic feedback which in their simulation is essential for quenching. In our simulation, the quenched galaxies are not solely quenched by the AGN feedback alone, as shown before, and thus we also find them in a different kind of environment. This environment in fact plays an important role, as we will see in more detail in Sec.~\ref{sec:env}. 
\begin{figure*}
  \begin{center}
    \includegraphics[width=.9\textwidth]{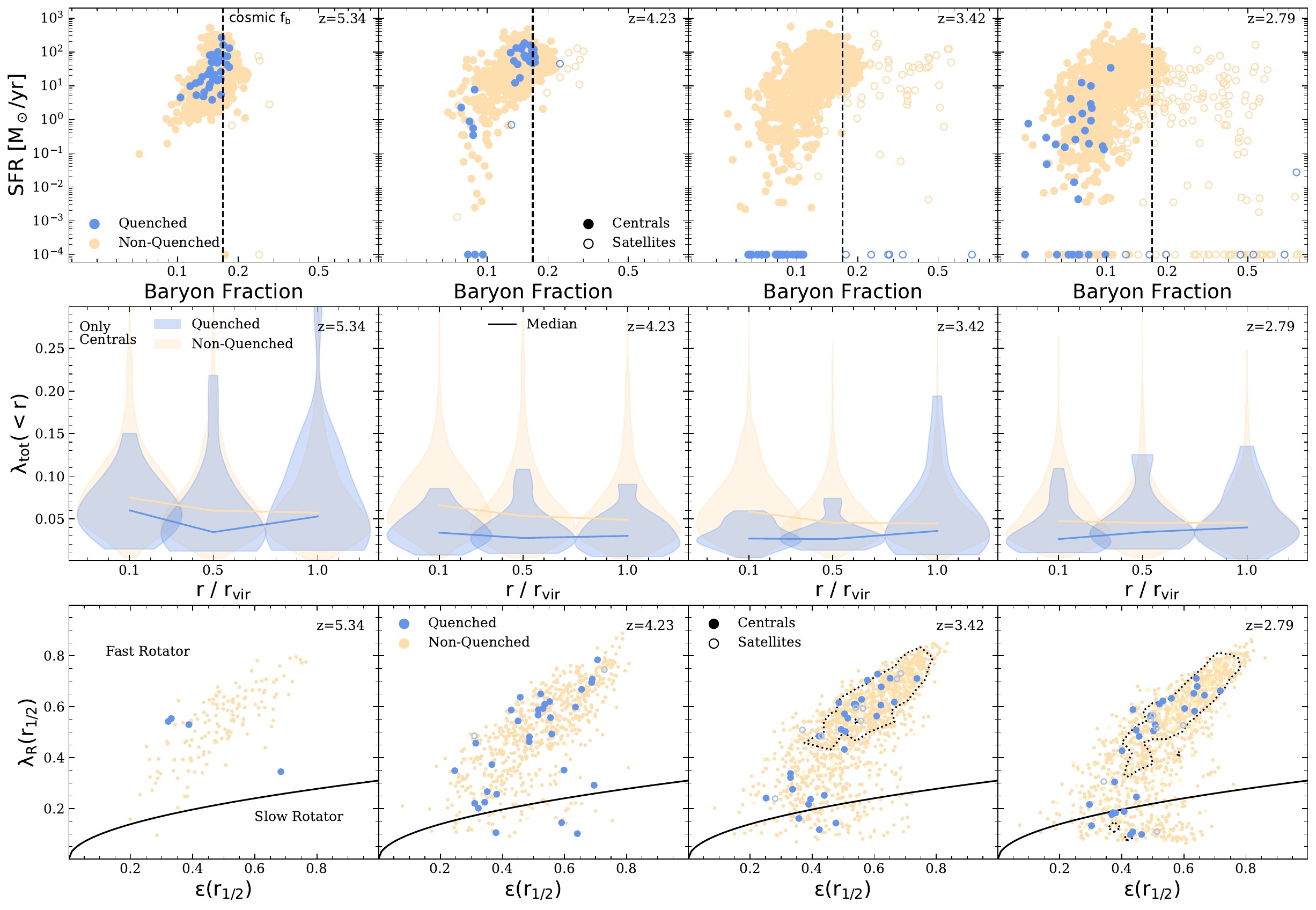}
    \caption{Various galaxy properties of the traced galaxy population going from high ($z=5.34$, left column) to intermediate redshift ($z=2.79$, right column). The quenched sample is defined by its SFR at $z=3.42$. \textit{Top row:} Star formation rate of traced quenched (blue) and non-quenched (sand) satellites (open symbols) and centrals (filled symbols) as a function of the baryon fraction (for centrals defined as within the virial radius). The cosmic baryon fraction is indicated by vertical dashed line. \textit{Middle row:} Distribution of total spin within different radii for the quenched (blue) and non-quenched (sand) centrals. Colored lines give the median. \textit{Bottom row:} $\lambda_R$ versus $\epsilon$ for the stars within one stellar half-mass radius. The solid line defines the split between fast and slow rotators given by \citet{emsellem:2011}, with the fast rotators lying above the line. Dashed contours enclose $50\%$ of all galaxies in the sample. 
    }
  {\label{fig:barfrac}}
\end{center}
\end{figure*}

We begin with the question whether the quenched galaxies have enhanced baryon fractions or even lower baryon fractions compared to their halo mass, something that is currently debated in the literature for most of the massive galaxies observed at high redshifts \citep[e.g.,][]{boylan:2023,harikane:2023_spec}.
The top row of Fig.~\ref{fig:barfrac} shows the baryon fraction $f_\mathrm{b}$ and star formation rate of the quenched (blue) and non-quenched (sand) galaxies at different redshifts, filled symbols marking central galaxies and open symbols those that are satellite galaxies of a more massive galaxy. $f_\mathrm{b}$ is defined within the virial radius for centrals, and for all bound particles in the case of satellites. 

Generally, we find that galaxies at higher baryon fractions can reach higher star formation rates on average, independent of the redshift. However, overall a significant fraction of the progenitors of both non-quenched and quenched galaxies lie at the cosmic baryon fraction of $16.8\%$ within the simulation, marked by the dashed line. Central galaxies tend to have lower baryon fractions, while galaxies with larger baryon fractions than the cosmic one tend to be satellite galaxies. For those, this enhancement in baryon fraction is clearly driven by the onset of stripping of the dark matter in their outskirts in the larger potential of the host, and as such this behavior is expected.

For our quenched galaxies identified at $z=3.42$, we do not find that they are overly baryon dominated prior to quenching when tracing them back in time, they rather behave exactly the same as the non-quenched sample. We find a slight tendency for those galaxies where their star formation has already declined but is not yet fully quenched to have lower baryon fractions

After quenching at $z=3.42$, there is a very clear separation between quenched centrals and satellites. The centrals all lie far below the cosmic $f_\mathrm{b}$ as they have ejected practically all of their gas but not their DM. On the other hand, the satellites lie at or above the cosmic $f_\mathrm{b}$, indicating that they have had significant amounts of their DM stripped away through interactions with their host, a behavior that is also seen for the non-quenched sample. 

Going to $z=2.79$, we see that some of the quenched centrals actually rekindle their star formation even though they all remain noticeably below the cosmic $f_\mathrm{b}$. They have yet to accrete significant amounts of gas, but what they have is sufficient to slowly begin forming stars again. To conclude, we find that quenched galaxies lie in halos of high dark matter fraction as a result of the quenching process, as opposed to quenching occurring due to the galaxies residing in overly massive halos.

Given how strongly the quenching process changes the baryon fraction throughout the entire halo, it is interesting to ask if it also imprints in the kinematics. To answer this we consider the dimensionless total spin $\lambda_\mathrm{tot}$ within a radius $r$ as defined by \citet{bullock:2001}:
\begin{equation}
    \lambda_\mathrm{tot}(r) = \frac{J(r)}{\sqrt{2} M(r) V_\mathrm{circ}(r) r},
\end{equation}
where $J(r)$ is the total angular momentum within $r$, $M(r)$ the total mass within $r$, and $V_\mathrm{circ}(r) = \sqrt{G M(r) / r}$ is the circular velocity. This form is based on the initially introduced spin parameter by \citet{peebles:1969}. For a thorough study of the properties of $\lambda_\mathrm{tot}$ within the Magneticum simulations, see \citet{teklu:2015}.

The central row of Fig.~\ref{fig:barfrac} shows the total spin measured within different radii around the galaxies at different times. Again, the sample of galaxies that are identified to be quenched at $z=3.42$ is shown in blue, while the results for the total galaxy population are shown in sand. Since it is unclear what a halo total spin should be for a satellite galaxy, we only include those galaxies that are centrals in this figure.

At $z=5.34$ the total spin for the quenched galaxies is equal to the non-quenched sample in the outskirts, and largely similar in the center (see Fig.~\ref{fig:barfrac}). Both show slightly increasing rotation toward the center. However, it is already apparent that the peak total spin of the soon to be quenched galaxies is \textit{decreasing} toward the center. 

At $z=4.23$ (prior to quenching) this signal becomes significant, with noticeably reduced total spin throughout the entire galaxy, which is not the case for the comparison sample of non-quenched galaxies. This means that immediately prior to quenching the galaxies show surprisingly little rotation, even though they are undergoing a large starburst. This may indicate that the accretion of mass is happening isotropically, as the total halo spin is lowered on all radial scales prior to quenching. While tidal torque interactions usually enhance the halo spin \citep[e.g.,][]{peebles:1969}, the spin of the dark matter is usually removed through strong isotropic accretion or a major merger event on a counter-rotating orbit \citep{cervantes:2010}; however, we tested the latter and found that all of our quenched halos show multiple gas-rich merger events from multiple directions, and only two exhibit a major merger going from $z=5.34$ to $z=4.23$ (with zero major mergers going from $z=4.23$ to $z=3.42$). 

Indeed, post quenching ($z=3.42$ and $z=2.79$) the central region containing the galaxy shows the lowest amount of total spin, while in the outskirts we already find the spin starting to build up again to values comparable with the non-quenched sample. A positive total spin gradient with radius may then point toward the type of self-quenching burst found here, correlating with a preceding phase of isotropic infall. 

But does this large scale transformation of the galactic halo's spin also imprint on the stellar angular momentum within the central galaxy? To answer this we use the $\lambda_R$-parameter, a proxy for the projected angular momentum that can be measured from observations, introduced by \citet{emsellem:2007} and adapted for cosmological simulations assuming a constant mass-to-light ratio \citep[e.g.,][]{jesseit:2009,bois:2011}:

\begin{equation}
    \lambda_R = \frac{\sum_i R_i M_i |V_i|}{\sum_i R_i M_i \sqrt{|V_i|^2 + \sigma_i^2}},
\end{equation}
where the sums $\sum_i$ run over the Voronoi tessellation cells \citep{cappellari:2003}, $R_i$ is the projected distance of the Voronoi cell center, $M_i$ is the total mass, $V_i$ is the mean line-of-sight velocity, and $\sigma_i$ is the line-of-sight velocity dispersion, each as measured within the $i$th Voronoi cell. For a detailed study on the properties of $\lambda_R$ in the Magneticum simulations from $z=2$ to $z=0$, see \citet{schulze:2018}.

The bottom row of Fig.~\ref{fig:barfrac} shows $\lambda_R$ versus the ellipticity $\epsilon$ within the stellar half-mass radius for the sample of traced galaxies. Here we only plot a galaxy starting from when it reaches $M_*>3\times10^{10}M_\odot$ to ensure a good resolution of the central kinematics. The solid black line indicates the separation between fast and slow rotators, as determined by \citet{emsellem:2011}, with the fast rotators lying above the line.

For all galaxies (sand color), we find the vast majority to be fast rotating independent of the redshift, but we also see the onset of the formation of a slow rotating population towards lower redshifts, emerging at $z>4$. Interestingly, this does not correlate at all with our population of quiescent galaxies (blue color). In fact, the majority of the quenched galaxies identified at $z=3.42$ retain their rotation and remain fast rotators after quenching, clearly indicating that for the high-redshift galaxies the quenching process is not coupled to an increase in dispersion and a decrease in rotational velocity. This is in excellent agreement with observations by \citet{newman:2018}, who reported their four lensed quenched galaxies at $z=2-2.5$ to all be fast rotators.

The origin of the emerging slow rotating population is thus not the mechanism that quenches galaxies, as most of the slow rotating galaxies at these high redshifts are still star forming. This is different from observations of low-redshift slow rotating galaxies, which are usually quiescent elliptical galaxies \citep[e.g.,][]{emsellem:2007,falcon:2015,cortese:2016}, as the origin of these slow rotating galaxies is found to be mostly merger driven \citep[e.g.,][]{bois:2011,schulze:2018,lagos:2022}.
However, \citet{lagos:2022} also report from the EAGLE simulation a population of slow rotating galaxies at $z=0$ which have not been formed through merger events. They find these non-merger-quenched galaxies to be more compact and more recently quenchend, i.e., they retained star formation while possibly already being slowly rotating. This is in good agreement with the properties of our slow rotating galaxy population emerging between $z=4$ and $z=2$. However, the process that leads to the formation of these slow rotating galaxies at high redshifts is beyond the scope of this work and will be addressed in a future study.

We conclude for the population of quenched galaxies at high redshifts that on the one hand an isotropic infall of both gas and dark matter is essential, resulting in the total halo spin being significantly lowered compared to other galaxies. On the other hand, the isotropically infalling gas has time to redistribute its angular momentum into a disk-like configuration to make stars before the actual quenching occurs, as we find the quenched galaxies to be predominantly fast rotating with no clear offset to the total sample. Nevertheless, the imprint in the total halo spin clearly shows evidence for an isotropic accretion of matter onto the galaxy prior to quenching, and hints at the environment being an important aspect for the quenching to be successful which we explore in the following. 

\section{The role of environment}\label{sec:env}

Having noted the curious dip in total spin prior to quenching, we move here to analyze the surrounding environment of these galaxies. However, the multi-scalar nature of the cosmic web makes the task of linking the large-scale environment to galaxy properties an inherently difficult one, as a given galaxy might inhabit one environment (e.g., a 1-dimensional filament) at a certain scale, but a different one (e.g., a void) at another scale \citep{rieder:2013}. Modern structure finders like the Hessian-based NEXUS \citep{cautun:2013} respect this multi-scalar nature by applying a spherical filter of different radii and then combining structural elements identified at different scales. For flexible case-by-case galaxy environment studies, however, these algorithms remain cumbersome.

Instead, we consider the environment around the galaxies here in two ways. Firstly, by considering whether they live in generally over or underdense environments as given by the density contrast $\delta$. Secondly, by the geometric dimensionality parameter (hereafter referred to as $D$) as suggested by \citet{sarkar:2009}. 
As this section of the analysis focuses on the intermediate-to-large scale environments, all quantities computed in this section are in comoving units.
We start with the density contrast $\delta$, defined as
\begin{equation}
    \delta(<R)=\frac{\rho(<R)-\bar{\rho}}{\bar{\rho}}
\end{equation}
with $\rho(<R)$ the density within a sphere of radius $R$ around a galaxy and $\bar{\rho}$ the mean density at the current redshift. In order to be able to fully capture the morphology of the forming cosmic web around these galaxies, we choose an upper filter radius of $10\,$Mpc/h. This value is motivated by previous studies on the characteristic sizes of network components such as by \citet{aragon:2010}.

To better quantify the difference between the quenched and non-quenched centrals, we then define 
\begin{equation}
    \delta_\mathrm{\sigma,rel}(<R)=\frac{\delta(<R) - \bar{\delta}(<R)}{\mathrm{\sigma}_{\delta,R}}
\end{equation}
with $\bar{\delta}(<R)$ the mean density contrast within a sphere of radius $R$ for the galaxies in our sample and $\mathrm{\sigma}_{\delta,R}$ the standard deviation of the distribution of density contrasts.

By this definition $\delta_\mathrm{\sigma,rel}(<R)$ quantifies how strongly a galaxy's density contrast is an outlier compared to all the other galaxies within our sample. For example, $\delta_\mathrm{\sigma,rel}(<1\,\mathrm{Mpc})=+1$ would imply that the environment within $1\,\mathrm{Mpc}$~around the galaxy is denser compared to the total sample, being a $1\,\mathrm{\sigma}$ outlier. If we assume the overdensities of our sample are distributed normally this is equivalent to being denser than around~$84\%$ of other galaxies. 

\begin{figure}
  \begin{center}
    \includegraphics[width=.95\columnwidth]{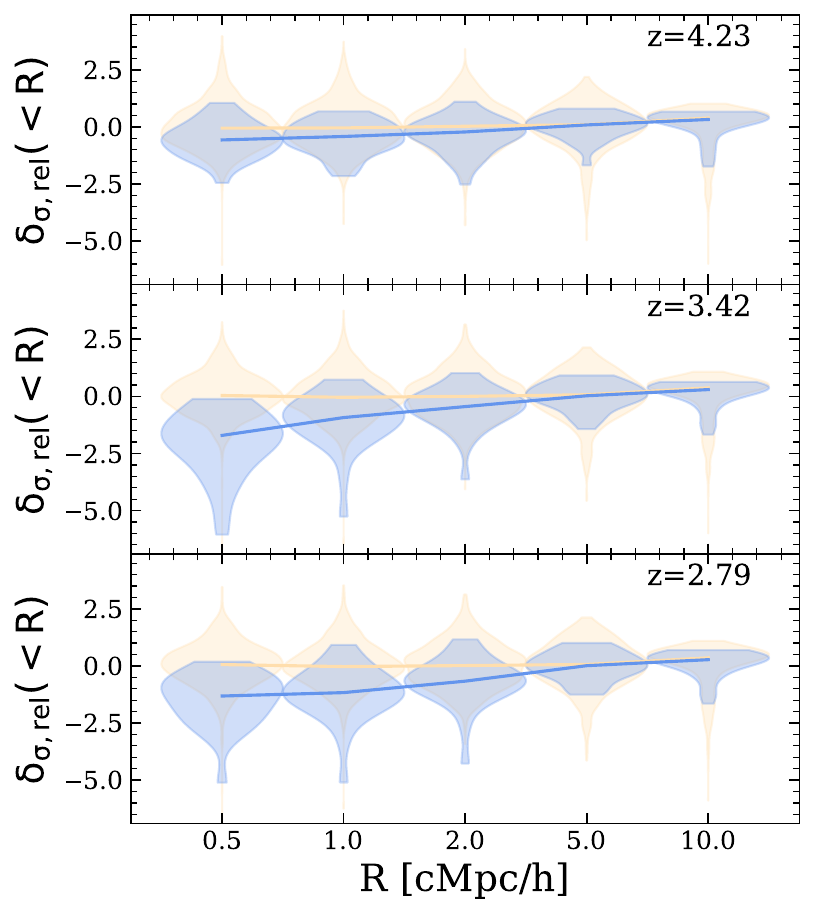}
    \caption{The deviation from the median density $\mathrm{\delta_{\sigma,rel}}$ within a sphere of radius $R$ for quenched (blue) and non-quenched (sand) centrals at three redshifts (\textit{top} to \textit{bottom}, $z=4.23$ to $z=2.79$). 
    }
  {\label{fig:env4}}
\end{center}
\end{figure}

Fig.~\ref{fig:env4} shows $\delta_\mathrm{\sigma,rel}(<R)$ for the quenched (blue) and non-quenched (sand) centrals for different radii. On the largest scales $R>5\,\mathrm{cMpc/h}$ the quenched galaxies lie in comparable environments to the rest of the sample, both before ($z=4.23$, top panel) and after quenching ($z=3.42$ and $z=2.79$, middle and bottom panels). 

However, the more locally we consider the environment around the galaxy, the more the quenched centrals deviate from those which do not quench. Interestingly, they inhabit comparably underdense regions (note that this does not mean underdense compared to the full simulation volume, but rather compared to galaxies of similar stellar mass). 

Prior to quenching they lie on average at $\delta_\mathrm{\sigma,rel}=-0.5$ for $R=0.5\,\mathrm{cMpc/h}$, while immediately post quenching this drops to $\delta_\mathrm{\sigma,rel}=-1.7$. This means that the galaxies which fully quench lie in more underdense regions than $96\%$ of galaxies of comparable stellar mass. Fascinatingly this signal is present even on larger scales of up to~$2\,\mathrm{cMpc/h}$, and also at $z=2.79$. 

We then move on to the dimensionality parameter $D$ \citep{sarkar:2009}. It is built on the simple assumption that the number of density tracers within a spherical filter scales with $N_{\mathrm{tracer}}\propto R_{\mathrm{filter}}^D$ in a D-dimensional spatial configuration. This means that $D=1$ is equivalent to the geometry of a filamentary structure, $D=2$ to a sheet and $D=3$ to a node (cluster) or antinode (void). It should be noted that intermediate values do not offer a clear interpretation as to the density structure, as the information is heavily convolved with the precise radial profile of the environment. This is why this parameter should be used only in joint analysis with the overdensity information itself, as is done within this section.

$D$ can be computed by 
\begin{equation}
    D\propto \frac{d\mathrm{log}(N)}{d\mathrm{log}(R_{\mathrm{filter}})}
\end{equation}
with $N$~the number of dark matter particles. We determine $D$ within 100 logarithmically spaced ($\Delta \log(R[\mathrm{cMpc/h}])=1.05$) spherical filters with radii ranging from $0.1\,\mathrm{cMpc/h}$ to $10\,\mathrm{cMpc/h}$. 

\begin{figure}
  \begin{center}
    \includegraphics[width=.95\columnwidth]{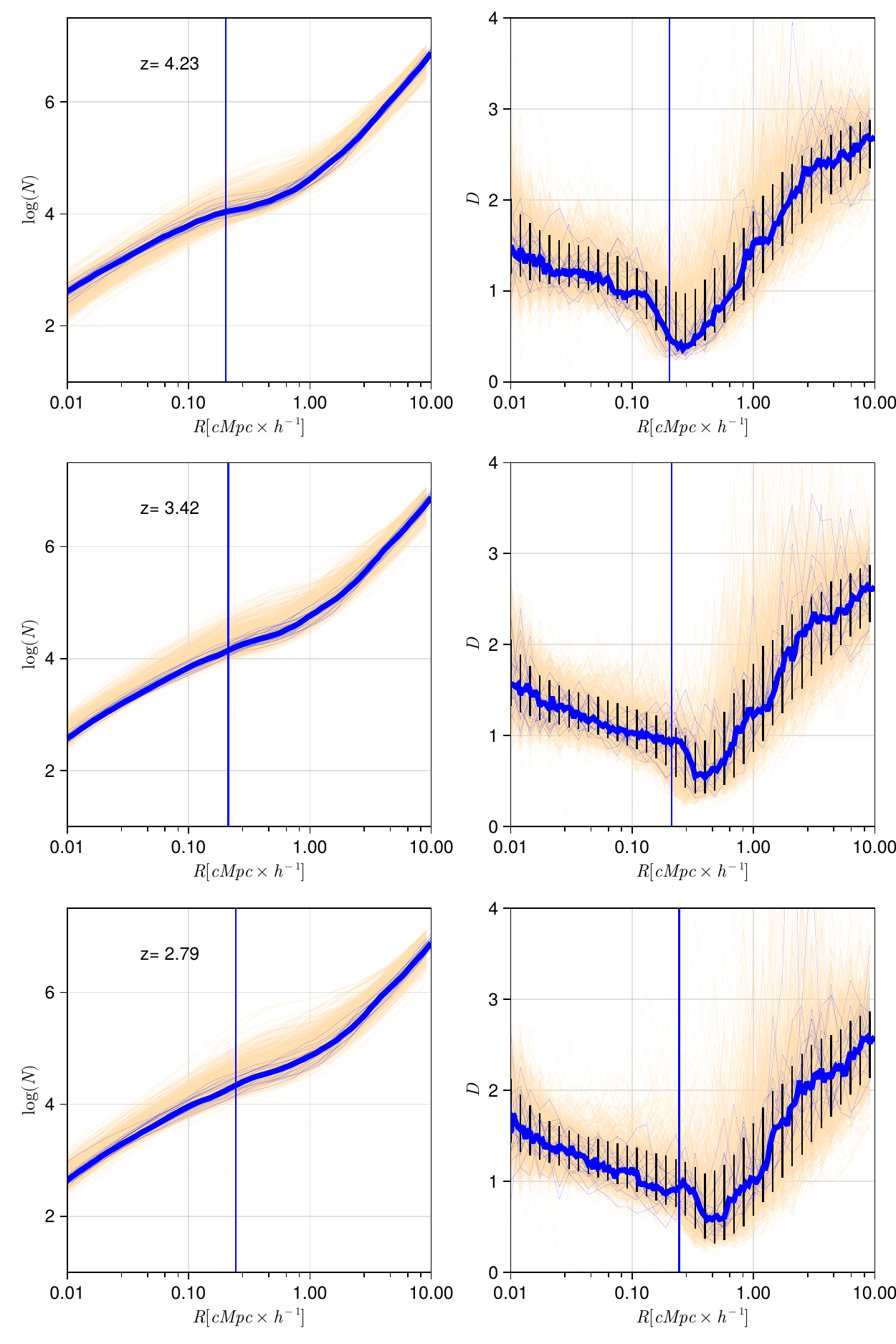}
    \caption{Number of dark matter particles $N$ (\textit{left panels}) and resulting dimensionality parameter $D$ (\textit{right panels}) as a function of radius around the quenched (blue lines) and non-quenched (sand lines) centrals, for three redshifts ($z=4.23$ to $z=2.79$, top to bottom). Right panels also show the $1\,\sigma$ bounds for the non-quenched centrals (vertical black lines) and median virial radii of the quenched centrals (vertical blue line). 
    }
  {\label{fig:env}}
\end{center}
\end{figure}

Fig.~\ref{fig:env} shows the distribution of dark matter around the centrals as a function of radius (left panels) as well as the resulting dimensionality $D$. At $z=4.23$ (top row) we find the quenched galaxies to not only exhibit an underdensity as expected from Fig.~\ref{fig:env4}, but find that $D$ drops noticeably below a value of~$1$ at around~$0.3\,\mathrm{cMpc/h}$. This implies a local shell of nearly null density (which shows up in the left panel as a region of nearly flat $N$). 

Comparing the profiles $N(R)$ between $z=4.23$ and $z=3.42$ for the quenched centrals, we find that they remain remarkably similar. Where the non-quenched galaxies have accreted additional dark matter towards their centers (thus generally reaching higher amounts of dark matter $N$ at $<1\,\mathrm{cMpc}$), the quenched halos have not. 

Combined with the results seen in Fig.~\ref{fig:env4}, we conclude that the galaxies which quench did not start out themselves being in severely underdense regions. Rather, they \textit{develop} into underdense regions coinciding with the time of quenching. This may be due either to other structures growing to become the locally dominant node or because there was always a region of low density nearby (or both, as for example seen in Fig.~\ref{fig:map}).

\section{Predictors For Quenching}\label{sec:disc}
\begin{figure*}
  \begin{center}
    \includegraphics[width=.9\textwidth]{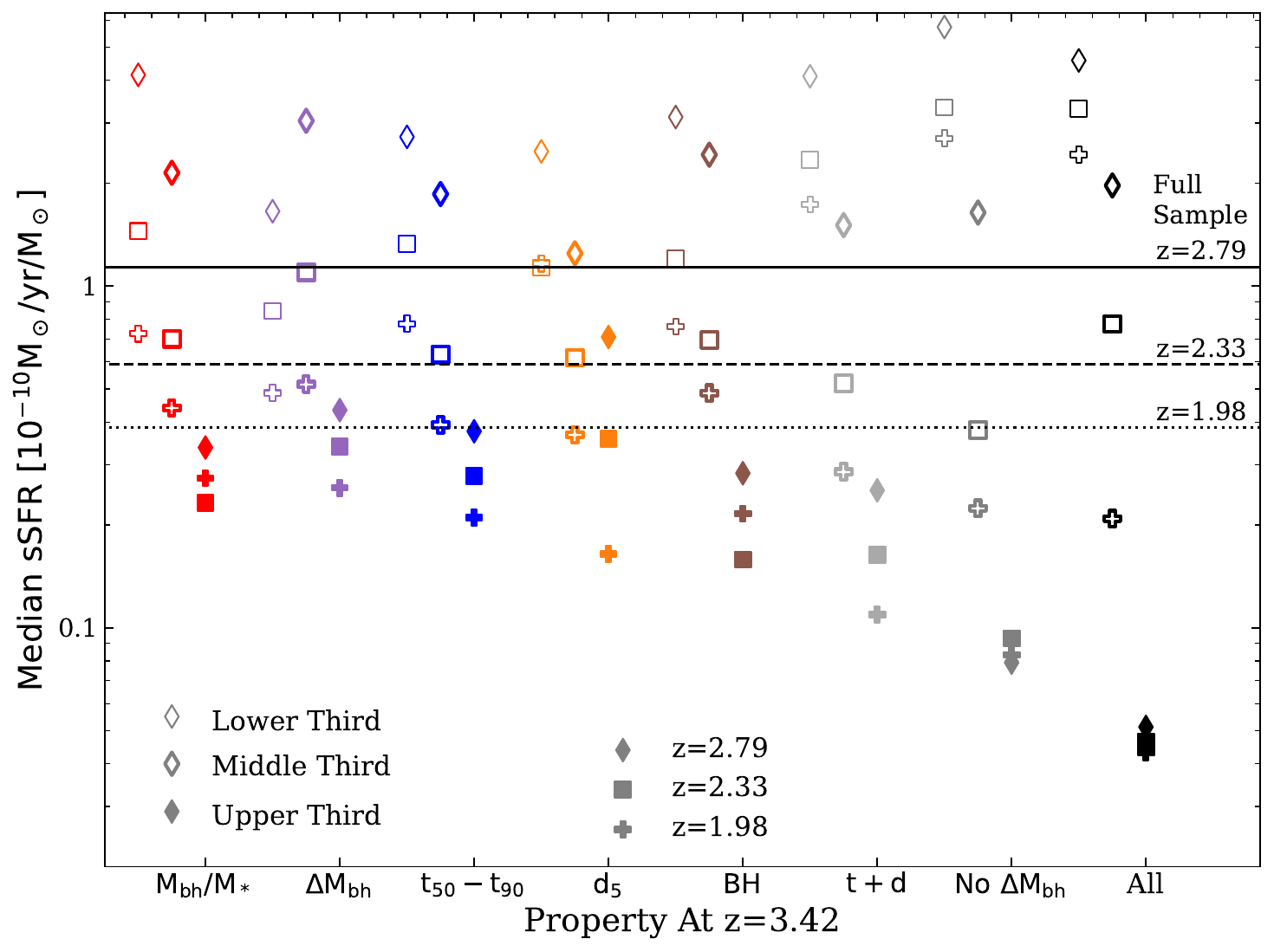}
    \caption{Median sSFR of $839$ centrals traced forward in time to $z=2.79$, $z=2.33$ and $z=1.98$ (diamond, square and plus markers). They are split into three groups based on whether they are in the lower, middle or upper third (open, semi-filled and filled symbols) of different parameters (colors) at $z=3.42$. Median sSFR for the full $839$~centrals are shown as horizontal black lines for the three redshifts. Combinations of parameters (brown, light gray, dark gray and black) show galaxies which lie in the lower/middle/upper third for all properties in the combination given on the x-axis, with the combinations being: ``BH'' for both black hole related properties, ``t+d'' for $t_\mathrm{50}-t_\mathrm{90}$ and $d_\mathrm{5}$, ``No $\Delta M_\mathrm{BH}$'' for all properties excluding the SMBH mass growth and ``All'' for all properties combined. 
    }
  {\label{fig:final}}
\end{center}
\end{figure*}
Throughout the prior sections we have shown the quenching process to correlate with a diverse set of galactic properties. We here quantify which parameters best predict future quenching, preferring observable quantities to best enable predictions for observed galaxies. 

Firstly, as discussed in Sec.~\ref{sec:quench}, quenched galaxies possess particularly massive black holes compared to their stellar mass (so have high $M_\mathrm{BH}/M_*$). We further find that they funnel significant amounts of gas towards their centers prior to quenching, which results in a large growth of their SMBH mass (high $\Delta M_\mathrm{BH}$). 
Aside from their SMBH properties, however, we find that the funneling of gas also results in a massive starburst (low $t_\mathrm{50}-t_\mathrm{90}$, see Sec.~\ref{subsec:rapid}). Finally, Sec.~\ref{sec:env} showed that they lie in generally underdense regions. 
\begin{table*}[!t]
    \centering
    \caption{Fraction (number) of central galaxies that lie in the upper third for a given property/combination of properties which become quenched at $z=2.79$, $z=1.98$ or at some point by $z=1.98$. The right-most column is the probability $P_\mathrm{rand}$ of randomly drawing the fraction of quenched galaxies given by the constraint of all properties excluding the recent mass growth of the SMBH (labeled ``Ex. $\Delta M_\mathrm{bh}$'')}
    \label{tab:frac_quench}
    \begin{tabular}{|r|ccccccccc|c|}
    \hline
    {Constraint} & {None} & {$M_\mathrm{bh}/M_*$} & {$\Delta M_\mathrm{bh}$} & {$t_\mathrm{50}-t_\mathrm{90}$} & {$d_\mathrm{5}$} & {BH} & {t+d} & {$\mathrm{Ex.}\ \Delta M_\mathrm{bh}$} & {All} & {$P_\mathrm{rand}$} \\
    {N$_\mathrm{galaxies}$} & 839 & 280 & 280 & 280 & 280 & 170 & 102 & 52 & 32 & - \\
    \hline
     {} & \multicolumn{9}{c|}{Fraction $[\%]$ (Number) Which Fully Quench} & {}\\
    \hline
    At z=2.79 & 5.1 (43) & 8.2 (23) & 7.9 (22) & 8.2 (23) & 7.1 (20) & 9.4 (16) & 9.8 (10) & 17.3 (9) & 18.8 (6) & 7.6e-4 \\
    At z=1.98 & 6.8 (57) & 10 (28) & 11.1 (31) & 13.2 (37) & 13.6 (38) & 11.1 (19) & 18.6 (19) & 25 (13) & 28.1 (9) & 1.4e-5 \\
    By z=1.98 & 13.3 (112) & 19.6 (55) & 18.9 (53) & 23.2 (65) & 21.1 (59) & 21.8 (37) & 31.4 (32) & 44.2 (23) & 53.1 (17) & $<$1e-6 \\
    \hline
    \end{tabular}
\end{table*}

We thus have four parameters to predict the future quenching (or lack thereof) for a given galaxy at high redshift. These are the black hole properties, $M_\mathrm{BH}/M_*$ and $\Delta M_\mathrm{BH}$, the star formation history quantified by the timescale $t_\mathrm{50}-t_\mathrm{90}$ on which stars are formed, and finally the environment quantified by the parameter $d_\mathrm{5}$. 
As opposed to the tracers of the environment used in Sec.~\ref{sec:env} (measuring \emph{all} mass), we use here instead $d_\mathrm{5}$, which is the minimum radius required to contain within it five other galaxies of total mass $M_\mathrm{all}\geq1\times10^{11}M_\odot$, aside from the galaxy in question. This is a much more reasonable characterization in terms of observability. RK23 show this parameter to be correlated nicely with the final halo mass at $z=0$, and that the quenched galaxies are also predominantly underdense by this definition of environment. 

In the following, we will now quantify the importance of all four parameters with respect to a long lasting quenching of a galaxy at high redshifts.
We therefore take the sample of $1217$~central galaxies at $z=3.42$ with a stellar mass of $M_*>3\times10^{10}M_\odot$, and keep only the $839$~which remain centrals until $z=2$ for this comparison, to exclude for example galaxies that merge onto more massive galaxies.

For every property we split the sample into three nearly equal groups (of around $280$~galaxies each) going from galaxies with values which should result in the lowest chance of quenching to those with the highest. For example, $M_\mathrm{BH}/M_*$ goes from lowest to highest (as the quenched galaxies have massive black holes) while $t_\mathrm{50}-t_\mathrm{90}$ goes from highest to lowest values (as the quenched galaxies show more rapid bursts of star formation). 

Every group of galaxies has their median specific star formation rate determined for the three redshifts of $z=2.79$, $z=2.33$ and $z=1.98$ (equivalent to times of around $0.5$,~$1$ and $1.5\,\mathrm{Gyr}$ from when the properties are determined at $z=3.42$). Fig.~\ref{fig:final} shows the resulting median sSFR for the $839$~centrals, with individual properties (red to orange) and combinations of properties (brown to black). 

When using a single property to determine whether a central galaxy will be starforming or quenched within the next $0.5\,\mathrm{Gyr}$, we find the strongest dependence on $M_\mathrm{BH}/M_*$. The galaxies with the lowest mass ratio (red open diamond) exhibit a noticeably enhanced sSFR of around~$4.2\times10^{-10} /\mathrm{yr}$, while those with the highest ratio (red filled diamond) have only around~$3.3\times10^{-11} 
/\mathrm{yr}$ compared to the total samples' average of~$1.1\times10^{-10} /\mathrm{yr}$. 

Only marginally less significant is the correlation with $t_\mathrm{50}-t_\mathrm{90}$ (blue), where galaxies which exhibit more prolonged star formation histories will likely continue forming more stars, while those which recently underwent a rapid burst are prone to find strongly reduced star formation compared to the total sample. 

When determining the star formation up to $z=1.98$, the picture changes. Both tracers for the SMBH perform more poorly than the other tracers, and in particular the tracer for environment (orange markers) becomes the most significant. On longer timescales the future of a galaxy is therefore most determined not by its internal properties, but rather the surrounding matter which may fall in. Even if a galaxy becomes quenched for some time along the way, if there is sufficient surrounding gas it may well rekindle its star formation (a point which RK23 discuss in more depth). Conversely, the galaxy may itself house the ideal conditions for a star formation burst (little SMBH feedback, no other recent burst) but if there is insufficient gas to fuel it, the burst will not happen. 

For the combined properties (brown to black markers) we note first that the number of galaxies per group is reduced. For a given combination (for example $t_\mathrm{50}-t_\mathrm{90}$ and $d\mathrm{5}$) the ``upper third'' would be the subset of galaxies which lie in the highest third of values for both properties \textit{simultaneously} (so for example galaxies which have the fastest $t_\mathrm{50}-t_\mathrm{90}$ \textit{and} largest $d_\mathrm{5}$). 

Fascinatingly, we find that at all redshifts the combination of star formation history and environment ($t_\mathrm{50}-t_\mathrm{90}$ and $d\mathrm{5}$, light gray ``t+d'') more significantly splits up the sample than the combined SMBH properties ($M_\mathrm{BH}/M_*$ and $\Delta M_\mathrm{BH}$, brown ``BH''). 

The difference is most pronounced when predicting the star formation rate up to $1.5\,\mathrm{Gyr}$ in the future. Galaxies which have the most prolonged SFH and lie in the densest environments will retain a comparably high specific star formation rate of~$1.7\times10^{-10}/\mathrm{yr}$ (light gray open diamond), while galaxies which have undergone recent bursts of star formation and which lie in underdense regions will on average exhibit very little star formation (around~$1.1\times10^{-11}/\mathrm{yr}$). 

That is not to say that the SMBH does not impact the significance of the prediction. When we further add $M_\mathrm{BH}/M_*$ to the SFH and environmental tracers, the split up becomes even more noticeable. On shorter timescales of $0.5\,\mathrm{Gyr}$ in particular, isolated galaxies which host massive central black holes and had a rapid burst of star formation exhibit a reduced specific star formation rate by over an order of magnitude compared to the total sample ($8\times10^{-12}/\mathrm{yr}$ versus $1.7\times10^{10}/\mathrm{yr}$). 

Including also the recent growth of the SMBH mass, so all tracers considered here simultaneously (black points), produces the strongest prediction towards low star formation rates. This, however, also represents the most stringent constraint, with just $32$~of $839$~central galaxies lying in the highest third for black hole mass, growth, fastest star formation burst and most underdense environment at the same time. 

Finally, we consider also which properties predict the future complete shut-down of star formation the best (galaxy reaches sSFR$=0/\mathrm{yr}$). To do so, we determine the number and fraction of galaxies within the ``upper third' samples for each property (and combination of properties) which become quenched. We consider three times, namely if they are quenched at $z=2.79$, at $z=1.98$, or if they were ever quenched between $z=3.42$ and $z=1.98$. The results are shown in Tab.~\ref{tab:frac_quench}.

Generally we find comparable behavior to that seen in Fig.~\ref{fig:final}. If a property results in a more reduced median sSFR, it also results in a higher fraction of quenched galaxies. We find that if a galaxy lies in the upper third for either massive $M_\mathrm{BH}/M_*$ or rapid $t_\mathrm{50}-t_\mathrm{90}$ compared to galaxies of like stellar mass, the probability that it will fully quench its star formation within $0.5\,\mathrm{Gyr}$ is around $8\%$ (compared to a base probability of $5\%$ for all galaxies). 

Combinations of properties provide more stringent conditions but also significantly increase the probability of the galaxy quenching. The combination of the observable quantities, so if a galaxy has a particularly massive black hole, a rapid burst of star formation and lies in a comparatively underdense region, results in a $17\%$ probability that the galaxy will fully quench within the next $0.5\,\mathrm{Gyr}$ ($9$~of the $52$~galaxies fulfilling the combined constraints). This chance is further enhanced to $44\%$ ($23$~galaxies) that it will be quenched at some point within the next $1.5\,\mathrm{Gyr}$. To confirm that this result is not a statistical outlier, we determine the probability of randomly selecting a subsample of $52$~galaxies from the full sample of~$839$ and finding at least~$23$ galaxies which fully quench. From a million random draws, not a single one finds $23$~or more quenched galaxies.

\section{Conclusions}\label{sec:conc}

Massive quiescent galaxies at cosmic dawn pose an interesting challenge to our understanding of the processes which occur at high redshift to build and transform galaxies. They must have undergone significant accretion and star formation on short timescales to reach the observed stellar masses, while simultaneously quenching it fully by the time they are observed. To explore how exactly these galaxies form and quench, we have employed~Box3~uhr of the Magneticum Pathfinder suite of cosmological simulations, which we have shown to fulfills both the requirements of high resolution (to resolve early bursty star formation) as well as box size (to contain more significant overdensities which can collapse early). 

We find that quenched galaxies undergo a noticeable starburst prior to quenching, resulting in some of the most peaked star formation histories even compared to other massive galaxies at high redshift. The rapid growth required to trigger such a large starburst means that they start out as \textit{local} attractors, quickly rising to become some of the most massive galaxies. Importantly, the influx of matter occurs rather isotropically compared to other massive galaxies at the same redshifts, and not directional along big filaments which is more common at lower redshifts.

Simultaneously, the internal processes of stellar and AGN feedback triggered by the preceding generally isotropic influx of massive amounts of gas have served to eject most gas far outside of the galaxy at the moment of quenching. Important for a successful quenching is hereby that the galaxies were only \textit{local} and not \textit{global} nodes of the cosmic web, and in fact reside in a globally underdense region of the web. The surrounding relative underdensity equates to weak feeding filaments, which on one hand allow for the gas to be ejected farther out, while on the other hand are themselves more easily disrupted. With all of its own gas consumed or ejected and no new gas streaming in, the galaxy is quenched. Whether it remains quenched or rejuvenates is discussed in greater depth by RK23. However, here we have shown an underdense environment to be the best indicator for a long-term quenching to be successful.

To summarize, we conclude that massive quenched galaxies at high redshifts generally
\begin{itemize}
    \item lie in significant \textbf{underdensities} compared to galaxies of similar stellar mass.
    \item went from $50\%$ to $90\%$ of their stellar mass in a \textbf{fast starburst} usually in less than $200\,\mathrm{Myr}$, comparable to recent observations \citep[e.g.,][]{kakimoto23,carnall:2023onequenchie}.
    \item host \textbf{massive central black holes}, but not more excessive than other galaxies of the same mass. 
    \item have \textbf{enhanced [O/Fe]$\approx0.14$} compared to non-quenched galaxies of the same stellar mass.
    \item have similar kinematic properties than non-quenched galaxies, mostly being \textbf{fast rotators}, in agreement with observations \citep{newman:2018}.
    \item are surrounded by \textbf{metal-rich ejected gas}.
\end{itemize}

We also investigate which quantities are most indicative of the future evolution of a given galaxy, especially with regards to observable quantities. For predicting whether a massive galaxy will be quenched in the future, we find that 
\begin{itemize}
    \item on short timescales a high black hole to stellar mass ratio $M_\mathrm{BH}/M_*$ is the best predictor, followed by a rapid starburst with short timescales of $t_\mathrm{50}-t_\mathrm{90}$, both of which are caused by the same infall of low angular momentum gas.
    \item on longer timescales it is instead the environment which becomes most significant, with less dense environments leading to the highest quenching likelihood due to the limited gas reservoir.
    \item on all timescales the combination of stellar and environmental properties is more relevant than the mass or recent growth of the SMBH.
    \item but including the requirement of a high SMBH mass further enhances the probability of a galaxy quenching within the next $1.5\,$Gyr to $44\%$. 
\end{itemize}

To conclude, we find evidence that quenching at high redshifts depends on more than just the AGN feedback, although the AGN feedback is a crucial component. In fact, the overall process of quenching is connected to a massive and rapid starburst event, caused by a fast isotropic accretion of gas that triggers both the starburst and also the growth of the SMBH. This ultimately leads to an alpha-enhancement of the stellar and gas metallicity through the short nature of the starburst and then full gas removal by the AGN feedback out to several virial radii in some cases. However, we find galaxies with similarly massive BHs, strong AGN feedback or short starbursts which do not quench. 

In fact, we conclude that quenching tends to depend on the infall of gas being rather isotropic and the environment being underdense. These conditions may be uniquely prevalent in the early universe, where strong feeding filaments have yet to form. \textit{We suggest that this rapid starburst quenching (RSQ) mechanism is the dominant mode of quenching down to a redshift of about $z=2$}, however, further studies on the efficiency of this type of quenching at different redshifts are required to confirm this in the future.

\section*{Acknowledgements}
We thank Deanne Fisher, Dominika Wylezalek, and Darren Croton for helpful discussions.
This work was supported by the Deutsche Forschungsgemeinschaft (DFG, German Research Foundation) under Germany's Excellence Strategy - EXC-2094 - 390783311.
LCK acknowledges support by the DFG project nr. 516355818. 
BS acknowledges supported by the grant agreements ANR-21-CE31-0019 / 490702358 from the French Agence Nationale de la Recherche / DFG for the LOCALIZATION project.
LMV acknowledges support by the German Academic Scholarship Foundation (Studienstiftung des deutschen Volkes) and the Marianne-Plehn-Program of the Elite Network of Bavaria.
KD and LK acknowledge support for the COMPLEX project from the European Research Council (ERC) under the European Union’s Horizon 2020 research and innovation program grant agreement ERC-2019-AdG 882679.
The {\it Magneticum} simulations were performed at the Leibniz-Rechenzentrum with CPU time assigned to the Project {\it pr83li}.  We are especially grateful for the support by M. Petkova through the Computational Center for Particle and Astrophysics (C2PAP).

\bibliography{biblio}

\begin{thebibliography}{110}
\expandafter\ifx\csname natexlab\endcsname\relax\def\natexlab#1{#1}\fi

\bibitem[{{Adams} {et~al.}(2023){Adams}, {Conselice}, {Ferreira}, {Austin},
  {Trussler}, {Juod{\v{z}}balis}, {Wilkins}, {Caruana}, {Dayal}, {Verma}, \&
  {Vijayan}}]{adams:2023}
{Adams}, N.~J., {Conselice}, C.~J., {Ferreira}, L., {et~al.} 2023, \mnras, 518,
  4755

\bibitem[{{Arag{\'o}n-Calvo} {et~al.}(2010){Arag{\'o}n-Calvo}, {van de
  Weygaert}, \& {Jones}}]{aragon:2010}
{Arag{\'o}n-Calvo}, M.~A., {van de Weygaert}, R., \& {Jones}, B. J.~T. 2010,
  \mnras, 408, 2163

\bibitem[{{Barnes}(1988)}]{barnes:1988}
{Barnes}, J.~E. 1988, \apj, 331, 699

\bibitem[{{Barnes}(1992)}]{barnes:1992}
{Barnes}, J.~E. 1992, \apj, 393, 484

\bibitem[{{Barnes} \& {Hernquist}(1996)}]{barnes:1996}
{Barnes}, J.~E. \& {Hernquist}, L. 1996, \apj, 471, 115

\bibitem[{{Barone} {et~al.}(2018){Barone}, {D'Eugenio}, {Colless}, {Scott},
  {van de Sande}, {Bland-Hawthorn}, {Brough}, {Bryant}, {Cortese}, {Croom},
  {Foster}, {Goodwin}, {Konstantopoulos}, {Lawrence}, {Lorente}, {Medling},
  {Owers}, \& {Richards}}]{barone:2018}
{Barone}, T.~M., {D'Eugenio}, F., {Colless}, M., {et~al.} 2018, \apj, 856, 64

\bibitem[{{Beck} {et~al.}(2016){Beck}, {Murante}, {Arth}, {Remus}, {Teklu},
  {Donnert}, {Planelles}, {Beck}, {F{\"o}rster}, {Imgrund}, {Dolag}, \&
  {Borgani}}]{beck:2015}
{Beck}, A.~M., {Murante}, G., {Arth}, A., {et~al.} 2016, \mnras, 455, 2110

\bibitem[{{Behroozi} \& {Silk}(2018)}]{behroozi:2018}
{Behroozi}, P. \& {Silk}, J. 2018, \mnras, 477, 5382

\bibitem[{{Belli} {et~al.}(2019){Belli}, {Newman}, \& {Ellis}}]{belli:2019}
{Belli}, S., {Newman}, A.~B., \& {Ellis}, R.~S. 2019, \apj, 874, 17

\bibitem[{{Belli} {et~al.}(2023){Belli}, {Park}, {Davies}, {Mendel}, {Johnson},
  {Conroy}, {Benton}, {Bugiani}, {Emami}, {Leja}, {Li}, {Maheson}, {Mathews},
  {Naidu}, {Nelson}, {Tacchella}, {Terrazas}, \& {Weinberger}}]{belli:2023}
{Belli}, S., {Park}, M., {Davies}, R.~L., {et~al.} 2023, arXiv e-prints,
  arXiv:2308.05795

\bibitem[{{Birnboim} \& {Dekel}(2003)}]{birnboim:2003}
{Birnboim}, Y. \& {Dekel}, A. 2003, \mnras, 345, 349

\bibitem[{{Bois} {et~al.}(2011){Bois}, {Emsellem}, {Bournaud}, {Alatalo},
  {Blitz}, {Bureau}, {Cappellari}, {Davies}, {Davis}, {de Zeeuw}, {Duc},
  {Khochfar}, {Krajnovi{\'c}}, {Kuntschner}, {Lablanche}, {McDermid},
  {Morganti}, {Naab}, {Oosterloo}, {Sarzi}, {Scott}, {Serra}, {Weijmans}, \&
  {Young}}]{bois:2011}
{Bois}, M., {Emsellem}, E., {Bournaud}, F., {et~al.} 2011, \mnras, 416, 1654

\bibitem[{{Bouwens} {et~al.}(2020){Bouwens}, {Gonz{\'a}lez-L{\'o}pez},
  {Aravena}, {Decarli}, {Novak}, {Stefanon}, {Walter}, {Boogaard}, {Carilli},
  {Dudzevi{\v{c}}i{\={u}}t{\.{e}}}, {Smail}, {Daddi}, {da Cunha}, {Ivison},
  {Nanayakkara}, {Cortes}, {Cox}, {Inami}, {Oesch}, {Popping}, {Riechers}, {van
  der Werf}, {Weiss}, {Fudamoto}, \& {Wagg}}]{bouwens:2020}
{Bouwens}, R., {Gonz{\'a}lez-L{\'o}pez}, J., {Aravena}, M., {et~al.} 2020,
  \apj, 902, 112

\bibitem[{{Boylan-Kolchin}(2023)}]{boylan:2023}
{Boylan-Kolchin}, M. 2023, Nature Astronomy, 7, 731

\bibitem[{{Bullock} {et~al.}(2001){Bullock}, {Dekel}, {Kolatt}, {Kravtsov},
  {Klypin}, {Porciani}, \& {Primack}}]{bullock:2001}
{Bullock}, J.~S., {Dekel}, A., {Kolatt}, T.~S., {et~al.} 2001, \apj, 555, 240

\bibitem[{{Cappellari} \& {Copin}(2003)}]{cappellari:2003}
{Cappellari}, M. \& {Copin}, Y. 2003, \mnras, 342, 345

\bibitem[{{Carnall} {et~al.}(2023{\natexlab{a}}){Carnall}, {McLeod}, {McLure},
  {Dunlop}, {Begley}, {Cullen}, {Donnan}, {Hamadouche}, {Jewell}, {Jones},
  {Pollock}, \& {Wild}}]{carnall:2023}
{Carnall}, A.~C., {McLeod}, D.~J., {McLure}, R.~J., {et~al.}
  2023{\natexlab{a}}, \mnras, 520, 3974

\bibitem[{{Carnall} {et~al.}(2018){Carnall}, {McLure}, {Dunlop}, \&
  {Dav{\'e}}}]{carnall:2018}
{Carnall}, A.~C., {McLure}, R.~J., {Dunlop}, J.~S., \& {Dav{\'e}}, R. 2018,
  \mnras, 480, 4379

\bibitem[{{Carnall} {et~al.}(2023{\natexlab{b}}){Carnall}, {McLure}, {Dunlop},
  {McLeod}, {Wild}, {Cullen}, {Magee}, {Begley}, {Cimatti}, {Donnan},
  {Hamadouche}, {Jewell}, \& {Walker}}]{carnall:2023onequenchie}
{Carnall}, A.~C., {McLure}, R.~J., {Dunlop}, J.~S., {et~al.}
  2023{\natexlab{b}}, \nat, 619, 716

\bibitem[{{Cautun} {et~al.}(2013){Cautun}, {van de Weygaert}, \&
  {Jones}}]{cautun:2013}
{Cautun}, M., {van de Weygaert}, R., \& {Jones}, B. J.~T. 2013, \mnras, 429,
  1286

\bibitem[{{Cervantes-Sodi} {et~al.}(2010){Cervantes-Sodi}, {Hernandez}, \&
  {Park}}]{cervantes:2010}
{Cervantes-Sodi}, B., {Hernandez}, X., \& {Park}, C. 2010, \mnras, 402, 1807

\bibitem[{{Combes}(2017)}]{combes:2017}
{Combes}, F. 2017, Frontiers in Astronomy and Space Sciences, 4, 10

\bibitem[{{Comer{\'o}n} {et~al.}(2023){Comer{\'o}n}, {Trujillo}, {Cappellari},
  {Buitrago}, {Gardu{\~n}o}, {Zaragoza-Cardiel}, {Zinchenko}, {Lara-L{\'o}pez},
  {Ferr{\'e}-Mateu}, \& {Dib}}]{comeron:2023}
{Comer{\'o}n}, S., {Trujillo}, I., {Cappellari}, M., {et~al.} 2023, \aap, 675,
  A143

\bibitem[{{Cortese} {et~al.}(2016){Cortese}, {Fogarty}, {Bekki}, {van de
  Sande}, {Couch}, {Catinella}, {Colless}, {Obreschkow}, {Taranu}, {Tescari},
  {Barat}, {Bland-Hawthorn}, {Bloom}, {Bryant}, {Cluver}, {Croom},
  {Drinkwater}, {d'Eugenio}, {Konstantopoulos}, {Lopez-Sanchez}, {Mahajan},
  {Scott}, {Tonini}, {Wong}, {Allen}, {Brough}, {Goodwin}, {Green}, {Ho},
  {Kelvin}, {Lawrence}, {Lorente}, {Medling}, {Owers}, {Richards}, {Sharp}, \&
  {Sweet}}]{cortese:2016}
{Cortese}, L., {Fogarty}, L.~M.~R., {Bekki}, K., {et~al.} 2016, \mnras, 463,
  170

\bibitem[{{Curtis-Lake} {et~al.}(2023){Curtis-Lake}, {Carniani}, {Cameron},
  {Charlot}, {Jakobsen}, {Maiolino}, {Bunker}, {Witstok}, {Smit}, {Chevallard},
  {Willott}, {Ferruit}, {Arribas}, {Bonaventura}, {Curti}, {D'Eugenio},
  {Franx}, {Giardino}, {Looser}, {L{\"u}tzgendorf}, {Maseda}, {Rawle}, {Rix},
  {Rodr{\'\i}guez del Pino}, {{\"U}bler}, {Sirianni}, {Dressler}, {Egami},
  {Eisenstein}, {Endsley}, {Hainline}, {Hausen}, {Johnson}, {Rieke},
  {Robertson}, {Shivaei}, {Stark}, {Tacchella}, {Williams}, {Willmer},
  {Bhatawdekar}, {Bowler}, {Boyett}, {Chen}, {de Graaff}, {Helton}, {Hviding},
  {Jones}, {Kumari}, {Lyu}, {Nelson}, {Perna}, {Sandles}, {Saxena}, {Suess},
  {Sun}, {Topping}, {Wallace}, \& {Whitler}}]{curtislake:2023}
{Curtis-Lake}, E., {Carniani}, S., {Cameron}, A., {et~al.} 2023, Nature
  Astronomy, 7, 622

\bibitem[{{Dekel} \& {Birnboim}(2006)}]{dekel:2006}
{Dekel}, A. \& {Birnboim}, Y. 2006, \mnras, 368, 2

\bibitem[{{Dekel} \& {Cox}(2006)}]{dekel:2006etgs}
{Dekel}, A. \& {Cox}, T.~J. 2006, \mnras, 370, 1445

\bibitem[{{Desprez} {et~al.}(2023){Desprez}, {Martis}, {Asada}, {Sawicki},
  {Willott}, {Muzzin}, {Abraham}, {Brada{\v{c}}}, {Brammer},
  {Estrada-Carpenter}, {Iyer}, {Matharu}, {Mowla}, {Noirot}, {Sarrouh},
  {Strait}, {Gledhill}, \& {Rihtar{\v{s}}i{\v{c}}}}]{desprez:2023}
{Desprez}, G., {Martis}, N.~S., {Asada}, Y., {et~al.} 2023, arXiv e-prints,
  arXiv:2310.03063

\bibitem[{{Ding} {et~al.}(2022){Ding}, {Onoue}, {Silverman}, {Matsuoka},
  {Izumi}, {Strauss}, {Jahnke}, {Phillips}, {Li}, {Volonteri}, {Haiman},
  {Taufik Andika}, {Aoki}, {Baba}, {Bieri}, {Bosman}, {Bottrell}, {Eilers},
  {Fujimoto}, {Habouzit}, {Imanishi}, {Inayoshi}, {Iwasawa}, {Kashikawa},
  {Kawaguchi}, {Kohno}, {Lee}, {Lupi}, {Lyu}, {Nagao}, {Overzier}, {Schindler},
  {Schramm}, {Shimasaku}, {Toba}, {Trakhtenbrot}, {Trebitsch}, {Treu},
  {Umehata}, {Venemans}, {Vestergaard}, {Walter}, {Wang}, \&
  {Yang}}]{ding:2022}
{Ding}, X., {Onoue}, M., {Silverman}, J.~D., {et~al.} 2022, arXiv e-prints,
  arXiv:2211.14329

\bibitem[{{Dolag} {et~al.}(2009){Dolag}, {Borgani}, {Murante}, \&
  {Springel}}]{dolag:2009}
{Dolag}, K., {Borgani}, S., {Murante}, G., \& {Springel}, V. 2009, \mnras, 399,
  497

\bibitem[{{Dolag} {et~al.}(2004){Dolag}, {Jubelgas}, {Springel}, {Borgani}, \&
  {Rasia}}]{dolag:2004}
{Dolag}, K., {Jubelgas}, M., {Springel}, V., {Borgani}, S., \& {Rasia}, E.
  2004, \apjl, 606, L97

\bibitem[{{Dolag} {et~al.}(2017){Dolag}, {Mevius}, \& {Remus}}]{dolag:2017}
{Dolag}, K., {Mevius}, E., \& {Remus}, R.-S. 2017, Galaxies, 5, 35

\bibitem[{{Dolag} {et~al.}(2005){Dolag}, {Vazza}, {Brunetti}, \&
  {Tormen}}]{dolag:2005}
{Dolag}, K., {Vazza}, F., {Brunetti}, G., \& {Tormen}, G. 2005, \mnras, 364,
  753

\bibitem[{{Dome} {et~al.}(2023){Dome}, {Tacchella}, {Fialkov}, {Dekel},
  {Ginzburg}, {Lapiner}, \& {Looser}}]{dome:2023}
{Dome}, T., {Tacchella}, S., {Fialkov}, A., {et~al.} 2023, arXiv e-prints,
  arXiv:2305.07066

\bibitem[{{Donnan} {et~al.}(2023){Donnan}, {McLeod}, {Dunlop}, {McLure},
  {Carnall}, {Begley}, {Cullen}, {Hamadouche}, {Bowler}, {Magee}, {McCracken},
  {Milvang-Jensen}, {Moneti}, \& {Targett}}]{donan:2023}
{Donnan}, C.~T., {McLeod}, D.~J., {Dunlop}, J.~S., {et~al.} 2023, \mnras, 518,
  6011

\bibitem[{{Donnert} {et~al.}(2013){Donnert}, {Dolag}, {Brunetti}, \&
  {Cassano}}]{donnert:2013}
{Donnert}, J., {Dolag}, K., {Brunetti}, G., \& {Cassano}, R. 2013, \mnras, 429,
  3564

\bibitem[{{D'Silva} {et~al.}(2023{\natexlab{a}}){D'Silva}, {Driver}, {Lagos},
  {Robotham}, {Bellstedt}, {Davies}, {Thorne}, {Bland-Hawthorn}, {Bravo},
  {Holwerda}, {Phillipps}, {Seymour}, {Siudek}, \& {Windhorst}}]{dsilva:2023}
{D'Silva}, J. C.~J., {Driver}, S.~P., {Lagos}, C. D.~P., {et~al.}
  2023{\natexlab{a}}, \mnras, 524, 1448

\bibitem[{{D'Silva} {et~al.}(2023{\natexlab{b}}){D'Silva}, {Lagos}, {Davies},
  {Lovell}, \& {Vijayan}}]{dsilva:2023_flares_sharks}
{D'Silva}, J. C.~J., {Lagos}, C. D.~P., {Davies}, L. J.~M., {Lovell}, C.~C., \&
  {Vijayan}, A.~P. 2023{\natexlab{b}}, \mnras, 518, 456

\bibitem[{{Dutton} {et~al.}(2016){Dutton}, {Macci{\`o}}, {Dekel}, {Wang},
  {Stinson}, {Obreja}, {Di Cintio}, {Brook}, {Buck}, \& {Kang}}]{dutton:2016}
{Dutton}, A.~A., {Macci{\`o}}, A.~V., {Dekel}, A., {et~al.} 2016, \mnras, 461,
  2658

\bibitem[{{Emsellem} {et~al.}(2011){Emsellem}, {Cappellari}, {Krajnovi{\'c}},
  {Alatalo}, {Blitz}, {Bois}, {Bournaud}, {Bureau}, {Davies}, {Davis}, {de
  Zeeuw}, {Khochfar}, {Kuntschner}, {Lablanche}, {McDermid}, {Morganti},
  {Naab}, {Oosterloo}, {Sarzi}, {Scott}, {Serra}, {van de Ven}, {Weijmans}, \&
  {Young}}]{emsellem:2011}
{Emsellem}, E., {Cappellari}, M., {Krajnovi{\'c}}, D., {et~al.} 2011, \mnras,
  414, 888

\bibitem[{{Emsellem} {et~al.}(2007){Emsellem}, {Cappellari}, {Krajnovi{\'c}},
  {van de Ven}, {Bacon}, {Bureau}, {Davies}, {de Zeeuw}, {Falc{\'o}n-Barroso},
  {Kuntschner}, {McDermid}, {Peletier}, \& {Sarzi}}]{emsellem:2007}
{Emsellem}, E., {Cappellari}, M., {Krajnovi{\'c}}, D., {et~al.} 2007, \mnras,
  379, 401

\bibitem[{{Fabjan} {et~al.}(2010){Fabjan}, {Borgani}, {Tornatore}, {Saro},
  {Murante}, \& {Dolag}}]{fabjan:2010}
{Fabjan}, D., {Borgani}, S., {Tornatore}, L., {et~al.} 2010, \mnras, 401, 1670

\bibitem[{{Falc{\'o}n-Barroso} {et~al.}(2015){Falc{\'o}n-Barroso}, {Lyubenova},
  \& {van de Ven}}]{falcon:2015}
{Falc{\'o}n-Barroso}, J., {Lyubenova}, M., \& {van de Ven}, G. 2015, in Galaxy
  Masses as Constraints of Formation Models, ed. M.~{Cappellari} \&
  S.~{Courteau}, Vol. 311, 78--81

\bibitem[{{Fan} {et~al.}(2023){Fan}, {Ba{\~n}ados}, \& {Simcoe}}]{fan:2023}
{Fan}, X., {Ba{\~n}ados}, E., \& {Simcoe}, R.~A. 2023, \araa, 61, 373

\bibitem[{{Feldmann} \& {Mayer}(2015)}]{feldmann:2015}
{Feldmann}, R. \& {Mayer}, L. 2015, \mnras, 446, 1939

\bibitem[{{Finkelstein} {et~al.}(2015){Finkelstein}, {Ryan}, {Papovich},
  {Dickinson}, {Song}, {Somerville}, {Ferguson}, {Salmon}, {Giavalisco},
  {Koekemoer}, {Ashby}, {Behroozi}, {Castellano}, {Dunlop}, {Faber}, {Fazio},
  {Fontana}, {Grogin}, {Hathi}, {Jaacks}, {Kocevski}, {Livermore}, {McLure},
  {Merlin}, {Mobasher}, {Newman}, {Rafelski}, {Tilvi}, \&
  {Willner}}]{finkelstein:2015}
{Finkelstein}, S.~L., {Ryan}, Russell~E., J., {Papovich}, C., {et~al.} 2015,
  \apj, 810, 71

\bibitem[{{Franx} {et~al.}(2008){Franx}, {van Dokkum}, {F{\"o}rster Schreiber},
  {Wuyts}, {Labb{\'e}}, \& {Toft}}]{franx:2008}
{Franx}, M., {van Dokkum}, P.~G., {F{\"o}rster Schreiber}, N.~M., {et~al.}
  2008, \apj, 688, 770

\bibitem[{{Gabor} \& {Dav{\'e}}(2015)}]{gabor:2015}
{Gabor}, J.~M. \& {Dav{\'e}}, R. 2015, \mnras, 447, 374

\bibitem[{{Glazebrook} {et~al.}(2023){Glazebrook}, {Nanayakkara}, {Schreiber},
  {Lagos}, {Kawinwanichakij}, {Jacobs}, {Chittenden}, {Brammer}, {Kacprzak},
  {Labbe}, {Marchesini}, {Marsan}, {Oesch}, {Papovich}, {Remus}, {Tran},
  {Esdaile}, \& {Chandro Gomez}}]{glazebrook:2023}
{Glazebrook}, K., {Nanayakkara}, T., {Schreiber}, C., {et~al.} 2023, arXiv
  e-prints, arXiv:2308.05606

\bibitem[{{Haardt} \& {Madau}(2001)}]{haardt:2001}
{Haardt}, F. \& {Madau}, P. 2001, in Clusters of Galaxies and the High Redshift
  Universe Observed in X-rays, ed. D.~M. {Neumann} \& J.~T.~V. {Tran}, 64

\bibitem[{{Harikane} {et~al.}(2023{\natexlab{a}}){Harikane}, {Nakajima},
  {Ouchi}, {Umeda}, {Isobe}, {Ono}, {Xu}, \& {Zhang}}]{harikane:2023_spec}
{Harikane}, Y., {Nakajima}, K., {Ouchi}, M., {et~al.} 2023{\natexlab{a}}, arXiv
  e-prints, arXiv:2304.06658

\bibitem[{{Harikane} {et~al.}(2022){Harikane}, {Ono}, {Ouchi}, {Liu},
  {Sawicki}, {Shibuya}, {Behroozi}, {He}, {Shimasaku}, {Arnouts}, {Coupon},
  {Fujimoto}, {Gwyn}, {Huang}, {Inoue}, {Kashikawa}, {Komiyama}, {Matsuoka}, \&
  {Willott}}]{harikane:2022}
{Harikane}, Y., {Ono}, Y., {Ouchi}, M., {et~al.} 2022, \apjs, 259, 20

\bibitem[{{Harikane} {et~al.}(2023{\natexlab{b}}){Harikane}, {Ouchi}, {Oguri},
  {Ono}, {Nakajima}, {Isobe}, {Umeda}, {Mawatari}, \&
  {Zhang}}]{harikane:2023_phot}
{Harikane}, Y., {Ouchi}, M., {Oguri}, M., {et~al.} 2023{\natexlab{b}}, \apjs,
  265, 5

\bibitem[{{Harikane} {et~al.}(2023{\natexlab{c}}){Harikane}, {Zhang},
  {Nakajima}, {Ouchi}, {Isobe}, {Ono}, {Hatano}, {Xu}, \&
  {Umeda}}]{harikane:2023AGN}
{Harikane}, Y., {Zhang}, Y., {Nakajima}, K., {et~al.} 2023{\natexlab{c}}, arXiv
  e-prints, arXiv:2303.11946

\bibitem[{{Harris} {et~al.}(2020){Harris}, {Remus}, {Harris}, \&
  {Babyk}}]{harris:2020}
{Harris}, W.~E., {Remus}, R.-S., {Harris}, G. L.~H., \& {Babyk}, I.~V. 2020,
  \apj, 905, 28

\bibitem[{{Hartley} {et~al.}(2023){Hartley}, {Nelson}, {Suess}, {Garcia},
  {Park}, {Hernquist}, {Bezanson}, {Nevin}, {Pillepich}, {Schechter},
  {Terrazas}, {Torrey}, {Wellons}, {Whitaker}, \& {Williams}}]{hartley:2023}
{Hartley}, A.~I., {Nelson}, E.~J., {Suess}, K.~A., {et~al.} 2023, \mnras, 522,
  3138

\bibitem[{{Houston} {et~al.}(2023){Houston}, {Croton}, \&
  {Sinha}}]{houston:2023}
{Houston}, T., {Croton}, D.~J., \& {Sinha}, M. 2023, \mnras, 522, L11

\bibitem[{{Jesseit} {et~al.}(2009){Jesseit}, {Cappellari}, {Naab}, {Emsellem},
  \& {Burkert}}]{jesseit:2009}
{Jesseit}, R., {Cappellari}, M., {Naab}, T., {Emsellem}, E., \& {Burkert}, A.
  2009, \mnras, 397, 1202

\bibitem[{{Kakimoto} {et~al.}(2023){Kakimoto}, {Tanaka}, {Onodera},
  {Shimakawa}, {Wu}, {Gould}, {Ito}, {Jin}, {Kubo}, {Suzuki}, {Toft},
  {Valentino}, \& {Yabe}}]{kakimoto23}
{Kakimoto}, T., {Tanaka}, M., {Onodera}, M., {et~al.} 2023, arXiv e-prints,
  arXiv:2308.15011

\bibitem[{{Kannan} {et~al.}(2023){Kannan}, {Springel}, {Hernquist}, {Pakmor},
  {Delgado}, {Hadzhiyska}, {Hern{\'a}ndez-Aguayo}, {Barrera}, {Ferlito},
  {Bose}, {White}, {Frenk}, {Smith}, \& {Garaldi}}]{kannan:2023}
{Kannan}, R., {Springel}, V., {Hernquist}, L., {et~al.} 2023, \mnras, 524, 2594

\bibitem[{{Kimmig} {et~al.}(2023){Kimmig}, {Remus}, {Dolag}, \&
  {Biffi}}]{kimmig:2023}
{Kimmig}, L.~C., {Remus}, R.-S., {Dolag}, K., \& {Biffi}, V. 2023, \apj, 949,
  92

\bibitem[{{Komatsu} {et~al.}(2011){Komatsu}, {Smith}, {Dunkley}, {Bennett},
  {Gold}, {Hinshaw}, {Jarosik}, {Larson}, {Nolta}, \& {Page}}]{komatsu:2011}
{Komatsu}, E., {Smith}, K.~M., {Dunkley}, J., {et~al.} 2011, \apjs, 192, 18

\bibitem[{{Kormendy} \& {Ho}(2013)}]{kh13}
{Kormendy}, J. \& {Ho}, L.~C. 2013, \araa, 51, 511

\bibitem[{{Kurinchi-Vendhan} {et~al.}(2023){Kurinchi-Vendhan}, {Farcy},
  {Hirschmann}, \& {Valentino}}]{kurinchi:2023}
{Kurinchi-Vendhan}, S., {Farcy}, M., {Hirschmann}, M., \& {Valentino}, F. 2023,
  arXiv e-prints, arXiv:2310.03083

\bibitem[{{Labb{\'e}} {et~al.}(2023){Labb{\'e}}, {van Dokkum}, {Nelson},
  {Bezanson}, {Suess}, {Leja}, {Brammer}, {Whitaker}, {Mathews}, {Stefanon}, \&
  {Wang}}]{labbe:2023}
{Labb{\'e}}, I., {van Dokkum}, P., {Nelson}, E., {et~al.} 2023, \nat, 616, 266

\bibitem[{{Lagos} {et~al.}(2023){Lagos}, {Bravo}, {Tobar}, {Obreschkow},
  {Power}, {Robotham}, {Proctor}, {Hansen}, {Chandro-Gomez}, \&
  {Carrivick}}]{lagos:2023}
{Lagos}, C. D.~P., {Bravo}, M., {Tobar}, R., {et~al.} 2023, arXiv e-prints,
  arXiv:2309.02310

\bibitem[{{Lagos} {et~al.}(2022){Lagos}, {Emsellem}, {van de Sande},
  {Harborne}, {Cortese}, {Davison}, {Foster}, \& {Wright}}]{lagos:2022}
{Lagos}, C. d.~P., {Emsellem}, E., {van de Sande}, J., {et~al.} 2022, \mnras,
  509, 4372

\bibitem[{{Long} {et~al.}(2023){Long}, {Antwi-Danso}, {Lambrides}, {Lovell},
  {de la Vega}, {Valentino}, {Zavala}, {Casey}, {Wilkins}, {Yung}, {Arrabal
  Haro}, {Bagley}, {Bisigello}, {Chworowsky}, {Cooper}, {Cooper}, {Cooray},
  {Croton}, {Dickinson}, {Finkelstein}, {Franco}, {Gould}, {Hirschmann},
  {Hutchison}, {Kartaltepe}, {Kocevski}, {Koekemoer}, {Lucas}, {McKinney},
  {Papovich}, {Perez-Gonzalez}, {Pirzkal}, \& {Santini}}]{long:2023}
{Long}, A.~S., {Antwi-Danso}, J., {Lambrides}, E.~L., {et~al.} 2023, arXiv
  e-prints, arXiv:2305.04662

\bibitem[{{Lotz} {et~al.}(2021){Lotz}, {Dolag}, {Remus}, \&
  {Burkert}}]{lotz:2021}
{Lotz}, M., {Dolag}, K., {Remus}, R.-S., \& {Burkert}, A. 2021, \mnras, 506,
  4516

\bibitem[{{Lotz} {et~al.}(2019){Lotz}, {Remus}, {Dolag}, {Biviano}, \&
  {Burkert}}]{lotz:2019}
{Lotz}, M., {Remus}, R.-S., {Dolag}, K., {Biviano}, A., \& {Burkert}, A. 2019,
  \mnras, 488, 5370

\bibitem[{{Lovell} {et~al.}(2023{\natexlab{a}}){Lovell}, {Harrison},
  {Harikane}, {Tacchella}, \& {Wilkins}}]{lovell:2023}
{Lovell}, C.~C., {Harrison}, I., {Harikane}, Y., {Tacchella}, S., \& {Wilkins},
  S.~M. 2023{\natexlab{a}}, \mnras, 518, 2511

\bibitem[{{Lovell} {et~al.}(2023{\natexlab{b}}){Lovell}, {Roper}, {Vijayan},
  {Seeyave}, {Irodotou}, {Wilkins}, {Conselice}, {Fortuni}, {Kuusisto},
  {Merlin}, {Santini}, \& {Thomas}}]{lovell:2023b}
{Lovell}, C.~C., {Roper}, W., {Vijayan}, A.~P., {et~al.} 2023{\natexlab{b}},
  \mnras, 525, 5520

\bibitem[{{Lustig} {et~al.}(2023){Lustig}, {Strazzullo}, {Remus}, {D'Eugenio},
  {Daddi}, {Burkert}, {De Lucia}, {Delvecchio}, {Dolag}, {Fontanot}, {Gobat},
  {Mohr}, {Onodera}, {Pannella}, \& {Pillepich}}]{lustig:2023}
{Lustig}, P., {Strazzullo}, V., {Remus}, R.-S., {et~al.} 2023, \mnras, 518,
  5953

\bibitem[{{Madau} \& {Dickinson}(2014)}]{madau:2014}
{Madau}, P. \& {Dickinson}, M. 2014, \araa, 52, 415

\bibitem[{{Man} \& {Belli}(2018)}]{man:2018}
{Man}, A. \& {Belli}, S. 2018, Nature Astronomy, 2, 695

\bibitem[{{Merlin} {et~al.}(2012){Merlin}, {Chiosi}, {Piovan}, {Grassi},
  {Buonomo}, \& {La Barbera}}]{merlin:2012}
{Merlin}, E., {Chiosi}, C., {Piovan}, L., {et~al.} 2012, \mnras, 427, 1530

\bibitem[{{Nanayakkara} {et~al.}(2022){Nanayakkara}, {Glazebrook}, {Jacobs},
  {Schreiber}, {Brammer}, {Esdaile}, {Kacprzak}, {Labbe}, {Lagos},
  {Marchesini}, {Marsan}, {Nateghi}, {Oesch}, {Papovich}, {Remus}, \&
  {Tran}}]{nanayakkara:2022}
{Nanayakkara}, T., {Glazebrook}, K., {Jacobs}, C., {et~al.} 2022, arXiv
  e-prints, arXiv:2212.11638

\bibitem[{{Newman} {et~al.}(2018){Newman}, {Belli}, {Ellis}, \&
  {Patel}}]{newman:2018}
{Newman}, A.~B., {Belli}, S., {Ellis}, R.~S., \& {Patel}, S.~G. 2018, \apj,
  862, 126

\bibitem[{{Oesch} {et~al.}(2018){Oesch}, {Bouwens}, {Illingworth}, {Labb{\'e}},
  \& {Stefanon}}]{oesch:2018}
{Oesch}, P.~A., {Bouwens}, R.~J., {Illingworth}, G.~D., {Labb{\'e}}, I., \&
  {Stefanon}, M. 2018, \apj, 855, 105

\bibitem[{{Pacucci} {et~al.}(2023){Pacucci}, {Nguyen}, {Carniani}, {Maiolino},
  \& {Fan}}]{pacucci:2023}
{Pacucci}, F., {Nguyen}, B., {Carniani}, S., {Maiolino}, R., \& {Fan}, X. 2023,
  arXiv e-prints, arXiv:2308.12331

\bibitem[{{Park} {et~al.}(2023){Park}, {Belli}, {Conroy}, {Tacchella}, {Leja},
  {Cutler}, {Johnson}, {Nelson}, \& {Emami}}]{park:2023}
{Park}, M., {Belli}, S., {Conroy}, C., {et~al.} 2023, \apj, 953, 119

\bibitem[{{Peebles}(1969)}]{peebles:1969}
{Peebles}, P.~J.~E. 1969, \apj, 155, 393

\bibitem[{{Peirani} {et~al.}(2017){Peirani}, {Dubois}, {Volonteri},
  {Devriendt}, {Bundy}, {Silk}, {Pichon}, {Kaviraj}, {Gavazzi}, \&
  {Habouzit}}]{peirani:2017}
{Peirani}, S., {Dubois}, Y., {Volonteri}, M., {et~al.} 2017, \mnras, 472, 2153

\bibitem[{{Remus} {et~al.}(2023){Remus}, {Dolag}, \&
  {Dannerbauer}}]{remus:2023}
{Remus}, R.-S., {Dolag}, K., \& {Dannerbauer}, H. 2023, \apj, 950, 191

\bibitem[{{Rieder} {et~al.}(2013){Rieder}, {van de Weygaert}, {Cautun},
  {Beygu}, \& {Portegies Zwart}}]{rieder:2013}
{Rieder}, S., {van de Weygaert}, R., {Cautun}, M., {Beygu}, B., \& {Portegies
  Zwart}, S. 2013, \mnras, 435, 222

\bibitem[{{Sarkar} \& {Bharadwaj}(2009)}]{sarkar:2009}
{Sarkar}, P. \& {Bharadwaj}, S. 2009, \mnras, 394, L66

\bibitem[{{Schulze} {et~al.}(2018){Schulze}, {Remus}, {Dolag}, {Burkert},
  {Emsellem}, \& {van de Ven}}]{schulze:2018}
{Schulze}, F., {Remus}, R.-S., {Dolag}, K., {et~al.} 2018, \mnras, 480, 4636

\bibitem[{{Spiniello} {et~al.}(2023){Spiniello}, {D'Ago}, {Coccato}, {Hartke},
  {Tortora}, {Ferr{\'e}-Mateu}, {Pulsoni}, {Cappellari}, {Maksymowicz-Maciata},
  {Arnaboldi}, {Bevacqua}, {Gallazzi}, {Hunt}, {La Barbera},
  {Mart{\'\i}n-Navarro}, {Napolitano}, {Radovich}, {Scognamiglio}, {Spavone},
  \& {Zibetti}}]{spiniello:2023}
{Spiniello}, C., {D'Ago}, G., {Coccato}, L., {et~al.} 2023, arXiv e-prints,
  arXiv:2309.12966

\bibitem[{{Spiniello} {et~al.}(2021){Spiniello}, {Tortora}, {D'Ago}, {Coccato},
  {La Barbera}, {Ferr{\'e}-Mateu}, {Napolitano}, {Spavone}, {Scognamiglio},
  {Arnaboldi}, {Gallazzi}, {Hunt}, {Moehler}, {Radovich}, \&
  {Zibetti}}]{spiniello:2021}
{Spiniello}, C., {Tortora}, C., {D'Ago}, G., {et~al.} 2021, \aap, 646, A28

\bibitem[{{Springel}(2005)}]{springel:2005}
{Springel}, V. 2005, \mnras, 364, 1105

\bibitem[{{Springel} \& {Hernquist}(2003)}]{springel:2003}
{Springel}, V. \& {Hernquist}, L. 2003, \mnras, 339, 289

\bibitem[{{Springel} {et~al.}(2005){Springel}, {White}, {Jenkins}, {Frenk},
  {Yoshida}, {Gao}, {Navarro}, {Thacker}, {Croton}, {Helly}, {Peacock}, {Cole},
  {Thomas}, {Couchman}, {Evrard}, {Colberg}, \& {Pearce}}]{springel:2005b}
{Springel}, V., {White}, S. D.~M., {Jenkins}, A., {et~al.} 2005, \nat, 435, 629

\bibitem[{{Springel} {et~al.}(2001){Springel}, {White}, {Tormen}, \&
  {Kauffmann}}]{springel:2001}
{Springel}, V., {White}, S.~D.~M., {Tormen}, G., \& {Kauffmann}, G. 2001,
  \mnras, 328, 726

\bibitem[{{Steinborn} {et~al.}(2015){Steinborn}, {Dolag}, {Hirschmann},
  {Prieto}, \& {Remus}}]{steinborn:2015}
{Steinborn}, L.~K., {Dolag}, K., {Hirschmann}, M., {Prieto}, M.~A., \& {Remus},
  R.-S. 2015, \mnras, 448, 1504

\bibitem[{{Tacchella} {et~al.}(2022{\natexlab{a}}){Tacchella}, {Conroy},
  {Faber}, {Johnson}, {Leja}, {Barro}, {Cunningham}, {Deason}, {Guhathakurta},
  {Guo}, {Hernquist}, {Koo}, {McKinnon}, {Rockosi}, {Speagle}, {van Dokkum}, \&
  {Yesuf}}]{tacchella:2022}
{Tacchella}, S., {Conroy}, C., {Faber}, S.~M., {et~al.} 2022{\natexlab{a}},
  \apj, 926, 134

\bibitem[{{Tacchella} {et~al.}(2016){Tacchella}, {Dekel}, {Carollo},
  {Ceverino}, {DeGraf}, {Lapiner}, {Mandelker}, \& {Primack}}]{tacchella:2016}
{Tacchella}, S., {Dekel}, A., {Carollo}, C.~M., {et~al.} 2016, \mnras, 458, 242

\bibitem[{{Tacchella} {et~al.}(2022{\natexlab{b}}){Tacchella}, {Finkelstein},
  {Bagley}, {Dickinson}, {Ferguson}, {Giavalisco}, {Graziani}, {Grogin},
  {Hathi}, {Hutchison}, {Jung}, {Koekemoer}, {Larson}, {Papovich}, {Pirzkal},
  {Rojas-Ruiz}, {Song}, {Schneider}, {Somerville}, {Wilkins}, \&
  {Yung}}]{tacchella:2022_too_massive}
{Tacchella}, S., {Finkelstein}, S.~L., {Bagley}, M., {et~al.}
  2022{\natexlab{b}}, \apj, 927, 170

\bibitem[{{Tacchella} {et~al.}(2023){Tacchella}, {Johnson}, {Robertson},
  {Carniani}, {D'Eugenio}, {Kumari}, {Maiolino}, {Nelson}, {Suess},
  {{\"U}bler}, {Williams}, {Adebusola}, {Alberts}, {Arribas}, {Bhatawdekar},
  {Bonaventura}, {Bowler}, {Bunker}, {Cameron}, {Curti}, {Egami}, {Eisenstein},
  {Frye}, {Hainline}, {Helton}, {Ji}, {Looser}, {Lyu}, {Perna}, {Rawle},
  {Rieke}, {Rieke}, {Saxena}, {Sandles}, {Shivaei}, {Simmonds}, {Sun},
  {Willmer}, {Willott}, \& {Witstok}}]{tacchella:2023_threegal}
{Tacchella}, S., {Johnson}, B.~D., {Robertson}, B.~E., {et~al.} 2023, \mnras,
  522, 6236

\bibitem[{{Teklu} {et~al.}(2015){Teklu}, {Remus}, {Dolag}, {Beck}, {Burkert},
  {Schmidt}, {Schulze}, \& {Steinborn}}]{teklu:2015}
{Teklu}, A.~F., {Remus}, R.-S., {Dolag}, K., {et~al.} 2015, \apj, 812, 29

\bibitem[{{Tollet} {et~al.}(2019){Tollet}, {Cattaneo}, {Macci{\`o}}, {Dutton},
  \& {Kang}}]{tollet:2019}
{Tollet}, {\'E}., {Cattaneo}, A., {Macci{\`o}}, A.~V., {Dutton}, A.~A., \&
  {Kang}, X. 2019, \mnras, 485, 2511

\bibitem[{{Tornatore} {et~al.}(2007){Tornatore}, {Borgani}, {Dolag}, \&
  {Matteucci}}]{tornatore:2007}
{Tornatore}, L., {Borgani}, S., {Dolag}, K., \& {Matteucci}, F. 2007, \mnras,
  382, 1050

\bibitem[{{Tornatore} {et~al.}(2004){Tornatore}, {Borgani}, {Matteucci},
  {Recchi}, \& {Tozzi}}]{tornatore:2004}
{Tornatore}, L., {Borgani}, S., {Matteucci}, F., {Recchi}, S., \& {Tozzi}, P.
  2004, \mnras, 349, L19

\bibitem[{{Tozzi} {et~al.}(2023){Tozzi}, {Maiolino}, {Cresci}, {Piotrowska},
  {Belfiore}, {Curti}, {Mannucci}, \& {Marconi}}]{tozzi:2023}
{Tozzi}, G., {Maiolino}, R., {Cresci}, G., {et~al.} 2023, \mnras, 521, 1264

\bibitem[{{Vayner} {et~al.}(2023){Vayner}, {Zakamska}, {Ishikawa}, {Sankar},
  {Wylezalek}, {Rupke}, {Veilleux}, {Bertemes}, {Barrera-Ballesteros}, {Chen},
  {Diachenko}, {Goulding}, {Greene}, {Hainline}, {Hamann}, {Heckman},
  {Johnson}, {Lim}, {Liu}, {Lutz}, {Lutzgendorf}, {Mainieri}, {McCrory},
  {Murphree}, {Nesvadba}, {Ogle}, {Sturm}, \& {Whitesell}}]{vayner:2023}
{Vayner}, A., {Zakamska}, N.~L., {Ishikawa}, Y., {et~al.} 2023, arXiv e-prints,
  arXiv:2307.13751

\bibitem[{{Veilleux} {et~al.}(2023){Veilleux}, {Liu}, {Vayner}, {Wylezalek},
  {Rupke}, {Zakamska}, {Ishikawa}, {Bertemes}, {Barrera-Ballesteros}, {Chen},
  {Diachenko}, {Goulding}, {Greene}, {Hainline}, {Hamann}, {Heckman},
  {Johnson}, {Grace Lim}, {Lutz}, {L{\"u}tzgendorf}, {Mainieri}, {Maiolino},
  {McCrory}, {Murphree}, {Nesvadba}, {Ogle}, {Sankar}, {Sturm}, \&
  {Whitesell}}]{veilleux:2023}
{Veilleux}, S., {Liu}, W., {Vayner}, A., {et~al.} 2023, \apj, 953, 56

\bibitem[{Wang {et~al.}(2023)Wang, Leja, Atek, Labbe, Li, Bezanson, Brammer,
  Cutler, Dayal, Furtak, Greene, Kokorev, Pan, Price, Suess, Weaver, Whitaker,
  \& Williams}]{wang:2023}
Wang, B., Leja, J., Atek, H., {et~al.} 2023, Quantifying the Effects of Known
  Unknowns on Inferred High-redshift Galaxy Properties: Burstiness, the IMF,
  and Nebular Physics

\bibitem[{{Wiersma} {et~al.}(2009){Wiersma}, {Schaye}, \&
  {Smith}}]{wiersma:2009}
{Wiersma}, R.~P.~C., {Schaye}, J., \& {Smith}, B.~D. 2009, \mnras, 393, 99

\bibitem[{{Wild} {et~al.}(2016){Wild}, {Almaini}, {Dunlop}, {Simpson},
  {Rowlands}, {Bowler}, {Maltby}, \& {McLure}}]{wild:2016}
{Wild}, V., {Almaini}, O., {Dunlop}, J., {et~al.} 2016, \mnras, 463, 832

\bibitem[{{Wylezalek} {et~al.}(2020){Wylezalek}, {Flores}, {Zakamska},
  {Greene}, \& {Riffel}}]{wylezalek:2020}
{Wylezalek}, D., {Flores}, A.~M., {Zakamska}, N.~L., {Greene}, J.~E., \&
  {Riffel}, R.~A. 2020, \mnras, 492, 4680

\bibitem[{{Zinger} {et~al.}(2016){Zinger}, {Dekel}, {Kravtsov}, \&
  {Nagai}}]{zinger:2016}
{Zinger}, E., {Dekel}, A., {Kravtsov}, A.~V., \& {Nagai}, D. 2016, arXiv
  e-prints, arXiv:1610.02644

\end{thebibliography}
\bibliographystyle{aa}


\label{lastpage}
\end{document}